\documentclass[aps,prc,showpacs,showkeys,twoside,twocolumn,byrevtex,floatfix]{revtex4-2}

\usepackage[dvips]{graphicx} 
\usepackage{color}           

\usepackage{amsmath,amssymb,amsfonts}
\usepackage{bm}
\usepackage{mathtools}
\newcommand{\nc}{\newcommand}           
\nc{\vc}[1]     {\mbox{\boldmath $#1$}} 
\nc{\mapleft}[1]{                       
 \smash{\mathop{                      %
  \hbox to 0.90cm{\rightarrowfill} }\limits_{#1}}}

\nc{\beq}     {\begin{eqnarray}}
\nc{\eeq}    {\end{eqnarray}}
\nc{\bra}       {\langle}               
\nc{\ket}       {\rangle}               
\nc{\bras}[1]   {\langle#1|}            
\nc{\kets}[1]   {|#1\rangle}            
\nc{\del}       {\partial}              

\newcommand{\lw}[1]{\smash{\lower1.75ex\hbox{#1}}}

\nc{\red}[1]    {\textcolor{black}{#1}}  

\usepackage{hyperref}
\hypersetup{
  bookmarksnumbered,%
  colorlinks,%
  linkcolor=blue,%
  citecolor=blue,%
  urlcolor=blue,%
  plainpages=false,%
  pdfstartview=FitH
}

\nc{\mydraft}	{\setlength{\topmargin}{-1.5cm}}
\mydraft
\begin{document}

\title{
  Examination of the $\alpha$-cluster breaking in the four $0^+$ bands of $^{12}$C with the variation of multiple bases of the antisymmetrized molecular dynamics 
}

\author{Takayuki Myo}
\email{takayuki.myo@oit.ac.jp}
\affiliation{General Education, Faculty of Engineering, Osaka Institute of Technology, Osaka, Osaka 535-8585, Japan}
\affiliation{Research Center for Nuclear Physics (RCNP), Osaka University, Ibaraki, Osaka 567-0047, Japan}

\date{\today}

\begin{abstract}%
  I investigate $^{12}$C, particularly the four kinds of the $0^+$ bands with various types of the $3\alpha$ configurations.
  These states are obtained in the variation of the multiple bases of the antisymmetrized molecular dynamics,
  where the bases are optimized simultaneously in the variation of the total energy.
  In the $0^+_2$ Hoyle state and the $0^+_3$ linear-chain state,
  I confirm a mixture of the $\alpha$-cluster breaking of the $s$-wave configuration with contributions from the spin-orbit force,
  while the $0^+_4$ state exhibits a relatively pure $3\alpha$ cluster state characterized by a large radius.
  The $2^+_{2-4}$ and $4^+_{2-4}$ states also tend to be the pure $3\alpha$ cluster states.
  The monopole transitions between the $0^+_{2}$ and $0^+_{4}$ bands exhibit large values, suggesting
  breathing mode of the $3\alpha$ states in the $0^+_4$ band.
  This conclusion aligns with the predictions of the $3\alpha$ models with an $\alpha$ condensate and also with a neural network,
  although the order of the $0^+_3$ and $0^+_4$ bands is reversed in the present results due to the attraction of the spin-orbit force.
\end{abstract}

\maketitle

\section{Introduction}

Nuclear clustering is a fundamental property of nuclei \cite{ikeda68,horiuchi12,freer18}.
In the cluster states, some of the nucleons in a nucleus form clusters, such as the $\alpha$ particles, which develop spatially within the nucleus.
Cluster states are often observed near the threshold energy of the cluster emission due to the weak interaction between the isolated clusters.
This property is known as the ``threshold rule'' \cite{ikeda68}.

In this paper, I focus on the $3\alpha$ cluster states of $^{12}$C.
The $0^+_2$ state, known as the Hoyle state \cite{hoyle54}, is located slightly above the $3\alpha$ breakup threshold energy by 0.4 MeV
\cite{ajzenberg90,tunl,kelley17}.
This state is of great interest from the viewpoint of the $\alpha$ particle condensate nature \cite{tohsaki01,bo20}.
Many experiments have been performed on this state and its band members,
as well as in search of other candidates for cluster states in $^{12}$C \cite{freer07,hyldegaard10,itoh11,itoh13b,freer11,freer12,marin14,li22a,li22b}.
Recently, ab initio studies \cite{epelbaum12,carlson15,otsuka22} and algebraic approaches \cite{bijker20}
have been developed for the Hoyle state.

Theoretically, there have been many studies of $^{12}$C in the $3\alpha$ cluster model with different approaches 
\cite{bo20,kurokawa05,kurokawa07,kurokawa24,ohtsubo13,arai06,garrido15,funaki15,funaki16,bo16,imai19,ichikawa22,takemoto23,cheng25,ishikawa25},
where the $\alpha$ cluster is assumed to be an $s$-wave closed configuration.
Most studies predict four kinds of the $0^+$ bands with $2^+$ and $4^+$.
On the other hand, models without the $3\alpha$ cluster assumption obtain up to the third band \cite{kanada07,chernykh07,fukuoka13,suhara15}.
The order of the bands in terms of excitation energy also depends on the model;
In the $3\alpha$ cluster model, the $0^+_3$ band is considered to be a breathing mode of the $0^+_2$ band including the Hoyle state
due to spatial expansion \cite{bo20,bo16,ichikawa22,cheng25,kurokawa24,takemoto23}.
The $0^+_4$ band is a linear-chain structure of $3\alpha$ with a slight bending.
In the non-$3\alpha$ model, however, the $0^+_3$ band is a linear-chain state, and the breathing mode is not confirmed.
In this situation, examining the existence of the $0^+_4$ band in $^{12}$C with a non-$3\alpha$ cluster model
is important in relation to the possibility of the breathing mode of the $0^+_2$ band. 
This is the motivation of the present study of $^{12}$C.

For the theoretical model of $^{12}$C without cluster assumption, I employ the antisymmetrized molecular dynamics (AMD) \cite{kanada03}
that can describe both the shell-model states (mean-field states) and cluster states in a nucleus simultaneously.
These states can coexist in $^{12}$C, because the $3\alpha$ threshold is located at 7.7 MeV in the low excitation energy region \cite{ajzenberg90}.
Additionally, the spin-orbit force cannot be treated in the $3\alpha$ model due to the $s$-wave assumption of the $\alpha$ particle.
However, this force naively contributes to the ground state of $^{12}$C due to the $p_{3/2}$-closed configuration in a $jj$ coupling \cite{suhara13}.
It has also been suggested that the spin-orbit force is important for the order of the $0^+$ states in the excitation energy \cite{suhara15}.
In AMD, the spin-orbit force can be incorporated into the nuclear structure,
making this model appropriate for surveying of the fourth $0^+$ band in $^{12}$C.

In AMD, constructing the configurations to be superposed for a nucleus is an important task.
Recently, I and my collaborators have developed an effective scheme to optimize the multiple AMD configurations with respect to the energy variation of the total system,
in which we do not introduce any physical constraints to construct the configurations \cite{myo23b,myo25a,myo25b,myo26a}.
We have extended the method to generate the excited-state configurations by imposing orthogonality condition on the ground-state configurations.
In the previous works \cite{myo23b,myo25a,myo25b,myo26a}, we have applied this method to light nuclei, and
have discussed the emergence of the various cluster states alongside the shell-model states.
Recently, variation of multiple basis states with clusters have been developed combining with a control neural network
for light nuclei and hypernuclei \cite{cheng25,tian24,tian25}.

In AMD, the energy variation of the {single basis is referred to as the cooling method (the imaginary-time evolution).
I extend this method to the multiple AMD bases, referring to it as the multiple cooling or ``multicool method''.
In this study, I apply the multicool method to $^{12}$C for examining the $0^+$ band structures,
and discuss the transitions between the band states for monopole and quadrupole.
In particular, the monopole transition is beneficial for discussing the breathing mode of the cluster states in nuclei \cite{yamada08}.

In Sec.~\ref{sec:method}, I explain the energy variation in the multiple AMD configurations.
In Sec.~\ref{sec:result}, I discuss the results of $^{12}$C.
In Sec.~\ref{sec:summary}, I summarize this work.

\section{Theoretical methods}\label{sec:method}

\subsection{Variation of multiple bases of antisymmetrized molecular dynamics}\label{sec:AMD}

The framework of the multicool method is given in Refs.~\cite{myo23b,myo25a,myo25b,myo26a} and I briefly explain it in this paper.
Nuclear wave function of AMD, $\Phi_{\rm AMD}$, is defined as an antisymmetrized $A$-nucleon system \cite{kanada03}:
\begin{eqnarray}
  \begin{split}
\Phi_{\rm AMD}
&= {\cal A} \left\{ \phi_1,\cdots,\phi_A \right\} ,
\\
\phi_i(\bm{r})&=\left(\frac{2\nu}{\pi}\right)^{3/4} e^{-\nu(\bm{r}-\bm{Z}_i)^2} \chi_{\sigma,i} \chi_{\tau,i} ,
\\
\chi_{\sigma,i} &= \alpha^\uparrow_i \kets{\uparrow} + \alpha^\downarrow_i \kets{\downarrow}.
  \end{split}
\label{eq:AMD}
\end{eqnarray}
The single-nucleon wave function $\phi_i(\bm{r})$ with the particle index $i$ has a Gaussian wave packet with a range parameter $\nu$
and the centroid parameter $\bm{Z}_i$. 
The spin wave function $\chi_{\sigma,i}$ is a superposition of the up and down components with the amplitudes of $\alpha^{\uparrow/\downarrow}_i$.
The isospin component $\chi_{\tau,i}$ represents a proton or a neutron.
The variational parameters, $\{\bm{Z}_i,\alpha^{\uparrow/\downarrow}_i\}$, are complex numbers.
The real part of $\bm{Z}_i$ represents the mean position of a nucleon.

The energy variation of the single AMD basis state $\Phi_{\rm AMD}$ is given
in the following cooling equation (the imaginary-time evolution) \cite{kanada03}.
The minimization of the total intrinsic energy $E^\pm_{\rm AMD}$ is performed
with the parity projection operator $P^\pm$ and the Hamiltonian $H$, 
\begin{equation}
  \begin{split}
    \Phi^\pm_{\rm AMD}&= P^\pm \Phi_{\rm AMD}  \,, \quad
    E^\pm_{\rm AMD} = \dfrac{ \bra \Phi^\pm_{\rm AMD}|H| \Phi^\pm_{\rm AMD} \ket }{\bra \Phi^\pm_{\rm AMD}| \Phi^\pm_{\rm AMD} \ket},
    \\
    \dfrac{{\rm d} X_i}{{\rm d} t}&= \frac{\mu}{\hbar} \dfrac{\partial E^\pm_{\rm AMD}}{ \partial X_i^*},\quad \mbox{and c.c}.
  \end{split}
  \label{eq:cooling}
\end{equation}
Using an arbitrary negative number $\mu$, the parameters $\{X_i\}:=\{\bm{Z}_i,\alpha^{\uparrow/\downarrow}_i\}$ are determined.
After the variation, the angular-momentum projection with the operator $P^J_{MK}$ is performed as
\begin{eqnarray}
  \begin{split}
  \Psi^{J^\pm}_{MK,{\rm AMD}}
  &= P^J_{MK}P^{\pm} \Phi_{\rm AMD}.
  \end{split}
  \label{eq:projection}
\end{eqnarray}
where $J$, $M$, and $K$ are the total angular momentum, its $z$-component, and its component onto the intrinsic $z$ axis, respectively,

I superpose the AMD basis states having the individual sets of $\{X_i\}$ and with a number $N_{J^\pm}$. 
The total wave function $\Psi_{\rm t}^{J^\pm}$ is a superposition of the projected AMD basis states in Eq.~(\ref{eq:projection}),
denoted as $\Psi_n^{J^\pm}$ with the basis index $n$ and $J^\pm$ including the $K$-mixing as
\begin{eqnarray}
  \begin{split}
   \Psi_{\rm t}^{J^\pm}
&= \sum_{n=1}^{N_{J^\pm}} C_n^{J^\pm}\,  \Psi_n^{J^\pm} \,, \quad
   E_{\rm t}^{J^\pm} = \dfrac{ \bra \Psi_{\rm t}^{J^\pm} |H| \Psi_{\rm t}^{J^\pm} \ket }{\bra \Psi_{\rm t}^{J^\pm}| \Psi_{\rm t}^{J^\pm} \ket}. 
  \end{split}
   \label{eq:linear}
\end{eqnarray}
The variation of the total energy $E_{\rm t}^{J^\pm}$ leads to the generalized eigenvalue problem to obtain
$E_{\rm t}^{J^\pm}$ and $\{C_n^{J^\pm}\}$:
\begin{eqnarray}
  \begin{split}
   \sum_{n=1}^{N_{J^\pm}} \Bigl\{ \langle\Psi_{m}^{J^\pm} | H |\Psi_{n}^{J^\pm}\rangle - E_{\rm t}^{J^\pm} \langle\Psi_{m}^{J^\pm} | \Psi_{n}^{J^\pm} \rangle \Bigr\}\, C_n^{J^\pm} &= 0.
  \end{split}
   \label{eq:eigen}
\end{eqnarray}

I extend this energy variation to that in the multiple AMD basis states, which is a unique approach in the present study.
I define the wave function $\Phi$ as a linear combination of the intrinsic AMD basis states $\{\Phi_n\}$
corresponding to $\Phi^\pm_{\rm AMD}$ in Eq. (\ref{eq:cooling}), with a number $N_{\rm b}$ as 
\begin{equation}
  \begin{split}
   \Phi&= \sum_{n=1}^{N_{\rm b}} C_n\,  \Phi_n , \quad
   E    = \dfrac{ \bra \Phi |H| \Phi \ket }{\bra \Phi | \Phi \ket},
  \end{split}
  \label{eq:multi}
\end{equation}
where $E$ is the intrinsic total energy.
The parity projection is always performed and omit the notation of parity $(\pm)$ for simplicity.

In the previous studies \cite{myo23b,myo25a,myo25b,myo26a},
I extended the cooling equation for the multiple AMD basis states, named ``multicool method''.
In this paper, I explain the basic properties of the multicool method;
the multiple AMD configurations have the parameters of $\{X_{n,i}\}:=\{\bm{Z}_{n,i},\alpha^{\uparrow/\downarrow}_{n,i},C_n\}$
with the additional basis index $n$.
The cooling equation is extended from Eq. (\ref{eq:cooling}) as
\begin{equation}
  \begin{split}
   \dfrac{{\rm d} X_{n,i}}{{\rm d} t}&= \frac{\mu}{\hbar} \dfrac{\partial E}{ \partial X_{n,i}^*},\quad \mbox{and c.c}.
  \end{split}
  \label{eq:multi_eq}
\end{equation}
Using this equation, one can determine $\{X_{n,i}\}$ and obtain the ground-state configurations $\{\Phi_n\}$ in Eq.~(\ref{eq:multi}).
It is noted that the weights of the configurations $\Phi_n$ are typically at most around 30--40\% of the total wave function $\Phi$
\cite{myo23b,myo25a,myo25b,myo26a}, and are not very large.
This means that the superposition of the AMD configurations is important for expressing $\Phi$,
for example in the optimization of the intercluster relative wave function of a nucleus.
In addition, the rotational symmetry can also be restored in terms of the configurations with similar intrinsic shape with different rotations.

In the next step, to construct the configurations for the excited states, I introduce the orthogonality to the ground-state configurations $\{\Phi_n\}$.
In principle, achieving this orthogonality requires the angular-momentum projection of $\{\Phi_n\}$ \cite{kanada03},
but, this is computationally expensive.
As an alternative, I consider the several rotations of the configurations of $\{\Phi_n\}$.
In this study, I employ the two rotations; $(x,y,z)\to(z,x,y)$ and $(x,y,z)\to(y,z,x)$ as done in the previous studies \cite{myo23b,myo25a,myo25b,myo26a}.
Each rotation makes $N_{\rm b}$ configurations, and by adding the original $N_{\rm b}$ configurations,
$3N_{\rm b}$ configurations are considered in total for the ground state.
The ground-state configurations $\{\Phi_{c}\}$ are assigned with the index $c=1,\cdots,3N_{\rm b}$,
and I construct the excited-state configurations with the orthogonality to $\{\Phi_{c}\}$.
This method works successfully in the applications to light nuclei \cite{myo23b,myo25a,myo25b,myo26a}.

In this study, I utilize the projection operator method to obtain the excited-state configurations \cite{kukulin78};
the pseudopotential $V_\lambda$ is introduced in terms of $\{\Phi_{c}\}$ in the projection operator form with a positive strength $\lambda$.
I add $V_\lambda$ to the Hamiltonian and define $H_\lambda$ as follows
\begin{equation}
  \begin{split}
    H_\lambda &= H + V_\lambda,\quad
    V_\lambda = \lambda \sum_{c=1}^{3N_{\rm b}} \kets{\Phi_{c}}\bras{\Phi_{c}}.
  \end{split}
  \label{eq:PSE}
\end{equation}
The expectation value of $V_\lambda$ is positive, giving a repulsive effect on the total energy.
In the variation of energy with $H_\lambda$, the total wave function is obtained in order to reduce the repulsive contribution of $V_\lambda$,
that is to say, the overlap with $\{\Phi_{c}\}$.
If $\lambda$ is sufficiently large, the contribution of $V_\lambda$ becomes negligible,
and the total wave function becomes orthogonal to each configuration of $\{\Phi_{c}\}$ \cite{kukulin78}.
This method has been applied to the nuclear cluster systems in order to remove the Pauli-forbidden states \cite{myo14}.

Using $H_{\lambda}$, the total wave function $\Phi_{\lambda}$ and the total energy $E_{\lambda}$ are given as
\begin{equation}
  \begin{split}
    \Phi_{\lambda} &= \sum_{n=1}^{N_{\rm b}} C_{\lambda,n} \Phi_{\lambda,n},\quad
    E_{\lambda} = \dfrac{ \bra \Phi_{\lambda} |H_{\lambda}| \Phi_{\lambda} \ket }{\bra \Phi_{\lambda} | \Phi_{\lambda} \ket}.
  \end{split}
  \label{eq:multi2}
\end{equation}
The variation of $E_{\lambda}$ is performed using Eq. (\ref{eq:multi_eq}) and the configurations $\{\Phi_{\lambda,n}\}$ are determined.
After the variation, $E_{\lambda}$ is evaluated without the contribution of $V_\lambda$.

I perform the variation with a small $\lambda$ and repeat the calculation increasing $\lambda$.
With a small $\lambda$ value, $\Phi_{\lambda}$ is not yet orthogonal to $\{\Phi_c\}$,
but $\Phi_{\lambda}$ can be the excited state in a low excitation energy and $\{\Phi_{\lambda,n}\}$ can contribute to the low-lying states.
Hence I adopt $\{\Phi_{\lambda,n}\}$ with various values of $\lambda$
in the final wave function in Eq.~(\ref{eq:linear}).

I summarize the multicool method as follows:
\begin{enumerate}
\itemsep=2mm
\item[(i)]
  I prepare the $N_{\rm b}$ multiple AMD basis states and solve the multicool equation in Eq. (\ref{eq:multi_eq}).
  I obtain the ground-state configurations $\{\Phi_{n}\}$ and define $\{\Phi_{c}\}$ with $c=1,\cdots,3N_{\rm b}$ adding two kinds of rotations of $\{\Phi_n\}$.
  
\item[(ii)]
  I add the pseudopotential $V_\lambda$ to the Hamiltonian with various $\lambda$ in Eq.~(\ref{eq:PSE}).
  I solve the multicool equation with the basis number $N_{\rm b}$ and construct the configurations $\{\Phi_{\lambda,n}\}$ for the excited states.

\item[(iii)]
  I superpose $\{\Phi_n\}$ and $\{\Phi_{\lambda,n}\}$ with the angular-momentum and parity projections.
  I solve the eigenvalue problem of the Hamiltonian matrix in Eq. (\ref{eq:eigen}) and obtain the eigenstates for each $J^\pm$.
  Resonances are treated in the bound-state approximation.
\end{enumerate}

In the multicool calculation, the basis number $N_{\rm b}$ is determined to obtain the relevant configurations of the system
and is typically around 15--20 \cite{myo23b,myo25a,myo25b,myo26a}.
In the step (c), the total basis number is at most around 500 for $^{12}$C before the $K$-mixing.

Some of the specific properties of the multicool calculation are explained in Refs. \cite{myo23b,myo25a,myo25b,myo26a},
such as the resulting AMD configurations.
I often obtain the cluster configurations with the same cluster constituents, but with different intercluster distances.
This indicates the optimization of the relative wave function between clusters by superposing of the configurations.
In addition, the pseudopotential $V_\lambda$ efficiently constructs the appropriate multiple configurations for the excited states.
This is a key feature of the multicool method.

\subsection{Hamiltonian}\label{sec:ham}
I use the effective nuclear interactions consisting of the two-body central, spin-orbit ($LS$), and Coulomb forces.
I use the Volkov No. 2 for the central force \cite{volkov65} and the G3RS for the spin-orbit force \cite{tamagaki68,yamaguchi79}
according to the previous studies of $^{12}$C and other $p$-shell nuclei
\cite{bo20,funaki15,funaki16,bo16,imai19,itagaki05,suhara10,suhara13,suhara15,kanada03,myo23b,myo25a,myo26a}, in which the cluster states are treated.
I set the Majorana parameter $M$=0.625 (Wigner parameter $W=1-M$) and the Bartlett and Heisenberg parameters $B=H=0.125$,
and 1600 MeV of the spin-orbit strength.
This Hamiltonian is almost the same as that used in other calculations \cite{itagaki05,suhara13,suhara15,myo23b,myo25a,myo26a}.
For other interactions, the Gogny D1S force has been used for $^{12}$C with AMD \cite{isaka20,zhao21},
and the resulting energy spectrum is similar to that obtained in the present calculation.
Following the previous works of $^{12}$C \cite{bo16,itagaki05,suhara10,suhara13,suhara15,myo23b,myo25a,myo26a},
I set $\nu=0.235$ fm$^{-2}$ in Eq.~(\ref{eq:AMD}).

\begin{table}[t]
  \caption{
    Total energies and matter radii of the intrinsic states of $^{12}$C with positive and negative parities
    in the multicool calculation with $N_{\rm b}=16$, compared with the single AMD basis calculation.
  }\vspace*{0.1cm}
\label{tab:multi}
\renewcommand{\arraystretch}{1.5}
\begin{tabular}{p{2.2cm} rrp{0.2cm}rr }
\hline
              &  \multicolumn{2}{c}{Positive parity} && \multicolumn{2}{c}{Negative parity} \\ \cline{2-3} \cline{5-6}
              &   Single    & Multicool              &&   Single    & Multicool     \\
\hline
Energy (MeV)  &  $-74.63$   &  $-80.82$              &&   $-65.92$  &  $-72.49$     \\  
Radius (fm)   &     2.24    &     2.34               &&   2.52      &  2.64         \\
\hline
\end{tabular}
\end{table}

\section{Results}\label{sec:result}

\subsection{Energy variation} 

\begin{figure}[t]
\centering
\includegraphics[width=7.8cm,bb=0 0 360 252]{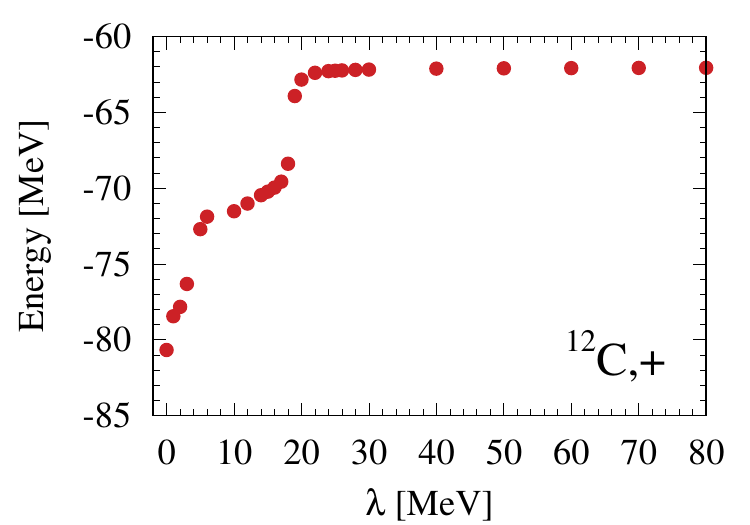}\\
\includegraphics[width=7.8cm,bb=0 0 360 252]{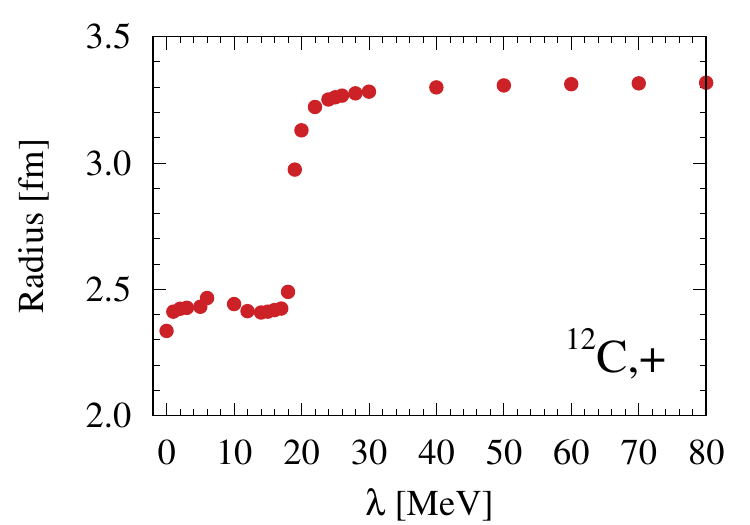}
\caption{
  Intrinsic energy (top) and matter radius (bottom) of $^{12}$C for a positive parity state
  with a basis number $N_{\rm b}=16$ in the multicool calculation.
  The strength $\lambda$ of the pseudopotential changes up to 80 MeV.}
\label{fig:ene_positive1}
\end{figure}

\begin{figure}[t]
\centering
\includegraphics[width=7.8cm,bb=0 0 360 252]{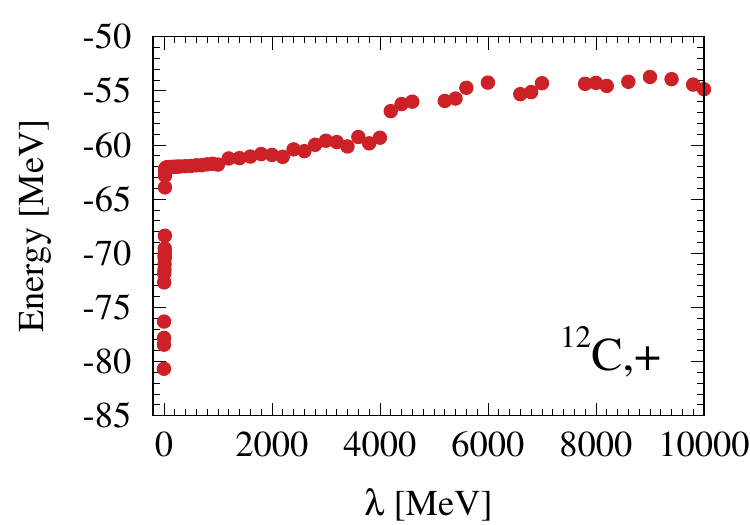}\\
\includegraphics[width=7.8cm,bb=0 0 360 252]{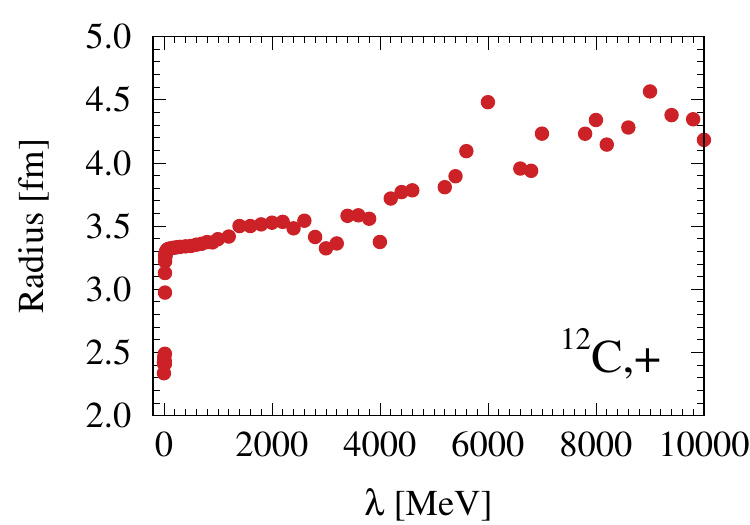}
\caption{
  The notation is the same as in Fig. \ref{fig:ene_positive1}, but the range of $\lambda$ extends up to 10000 MeV.}
\label{fig:ene_positive2}
\end{figure}

I perform the multicool calculation for $^{12}$C using the intrinsic AMD basis states with $N_{\rm b}=16$ in Eq.~(\ref{eq:multi}). 
The resulting energies for the positive and negative parities are summarized in Table \ref{tab:multi}.
I compare the results with those obtained using a single AMD basis state.
The energy gain due to the multiple bases is 6.2 (6.6) MeV for the positive (negative) parity state and these amounts are significant.
The radii increase slightly by around 0.1 fm in the both parities, indicating the mixing of the bases with large radii.
These results demonstrate the reliability of the multicool method in optimizing the nuclear multiple configurations.

Next, I construct the basis states for the excited states of $^{12}$C using the pseudopotential $V_\lambda$,
as defined in Eq. (\ref{eq:PSE}).
Figures~\ref{fig:ene_positive1} and \ref{fig:ene_positive2}
show the results for the total energies and matter radii of the positive parity state of $^{12}$C
as the strength $\lambda$ is varied.
The ranges of $\lambda$ are different in the two figures.
At $\lambda=0$, the calculation corresponds to the ground state, as shown in Table \ref{tab:multi}.

In Fig.~\ref{fig:ene_positive1}, by increasing $\lambda$, the energy starts to increase and
sudden changes in energy and radius occur around $\lambda=20$ MeV.
After this, the energy becomes stable at around $-62$ MeV, which corresponds to an excitation energy of 19 MeV.
The radius increases to around 3.3 fm, which is approximately 1 fm larger than the ground-state radius.
The stable behavior with respect to $\lambda$ indicates the construction of the excited state with a large radius.
In Fig.~\ref{fig:ene_positive2}, I increase $\lambda$ further to 10000 MeV, and observe that the energy gradually increases again,
reaching around $-55$ MeV at $\lambda=5000$ MeV. The radius becomes larger than 4 fm.
With a large $\lambda$, I can generate the excited states that are different from those obtained with a low $\lambda$ of around 30 MeV
in Fig.~\ref{fig:ene_positive1}.
I superpose the basis states obtained with specific $\lambda$ in the total wave function of $^{12}$C.
  In the individual basis states, their radii are at most around 6 fm,
  which is sufficient for describing the spatially extended $3\alpha$ cluster states located above the $3\alpha$ threshold energy
  \cite{funaki16,bo16,takemoto23}.
  In particular, in Ref. \cite{bo16}, the upper limit of the radius is introduced as 6 fm in the $3\alpha$ basis states of $^{12}$C.
  This value is determined to stabilize the energies of the $0^+_3$ and $0^+_4$ states avoiding the contamination from continuum states.
  I adopt the same criterion of the maximum radius of 6 fm for the multicool basis states of $^{12}$C.

\begin{figure}[t]
\centering
\includegraphics[width=8.6cm,bb=0 0 214 80]{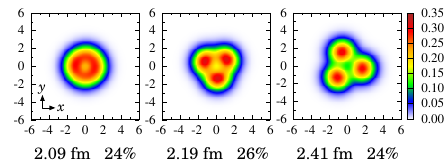}
\caption{
  Intrinsic density distributions of the representative configurations of $^{12}$C for the ground state. 
  Units of the densities with colorbar and the $x$-$y$ axes are in fm$^{-3}$ and in fm, respectively.
  The values below each panel represent the radii and the weights of the configurations in the total wave function.
}
\label{fig:density1}
\end{figure}
\begin{figure}[th]
\centering
\hspace*{-0.00cm}\includegraphics[width=8.60cm,bb=0 0 214 80]{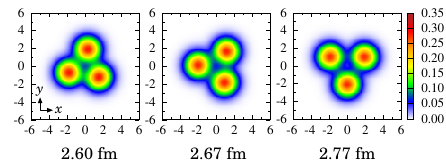}\\ 
\hspace*{-0.60cm}\includegraphics[width=7.95cm,bb=0 0 199 81]{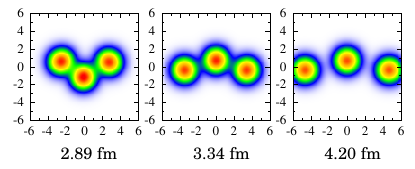}\\
\hspace*{-0.60cm}\includegraphics[width=7.95cm,bb=0 0 199 81]{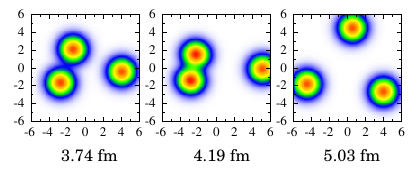}
\caption{
  Intrinsic density distributions of the representative configurations of $^{12}$C for the excited states
  using the pseudopotential $V_\lambda$
  with the strengths of $\lambda=50$ MeV (top three panels), 4000 MeV (middle three panels), and 8000 MeV (bottom three panels),
  corresponding to Figs. \ref{fig:ene_positive1} and \ref{fig:ene_positive2}.
  Units of the densities with colorbar and the $x$-$y$ axes are in fm$^{-3}$ and in fm, respectively.
  The radius is shown below each panel.
}
\label{fig:density2}
\end{figure}

I discuss the AMD configurations of $^{12}$C with positive parity obtained in the multicool calculation.
Figure~\ref{fig:density1} shows the intrinsic density distributions of the three representative configurations of
the intrinsic ground state of $^{12}$C.
In the figure, the longest distribution is set as the horizontal axis. 
The volume integral of each distribution yields a mass number.
The radii and the weights in the total wave function are also shown below each panel.
The density of the compact shell structure with a small radius of 2.1 fm is confirmed, 
as well as those of the compact $3\alpha$ clustering with larger radii.
The weights of the configurations are less than 40\%, which is typical behavior in the multicool calculation \cite{myo25a}.

I discuss the densities of the excited states shown in Fig.~\ref{fig:density2},
corresponding to Figs. \ref{fig:ene_positive1} and \ref{fig:ene_positive2}.
I show the three cases of the strengths; $\lambda=$ 50 MeV, 4000 MeV, and 8000 MeV in the pseudo potential $V_\lambda$. 
The distributions commonly show the $3\alpha$ clustering, but their geometrical configurations are different;
in the top three panels with $\lambda=50$ MeV,  triangle shapes are generated and the $\alpha$ particles are close and almost touching to each other. 
In the middle three panels with $\lambda=4000$ MeV, a linear-chain structure with a small bending are generated and the distances
between the neighboring $\alpha$ particles are changeable.
This indicates the effect of the generator coordinate of the $\alpha$-$\alpha$ relative distance.
In the bottom three panels with $\lambda=8000$ MeV,  one confirms the $^8$Be+$\alpha$ structure and also a large triangle configuration.
These configurations with different $\lambda$ values exhibit a variety of the $3\alpha$ structures,
and they become the sources for describing the excited states of $^{12}$C.

\begin{figure}[t]
\centering
\includegraphics[width=7.8cm,bb=0 0 360 252]{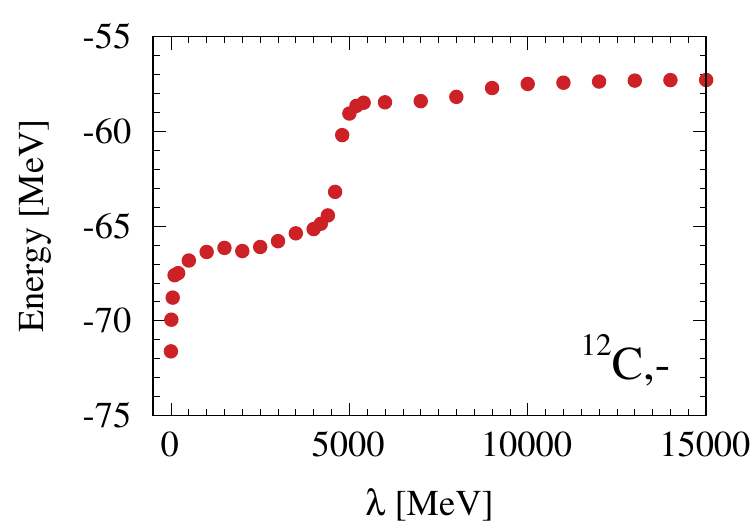}\\
\includegraphics[width=7.8cm,bb=0 0 360 252]{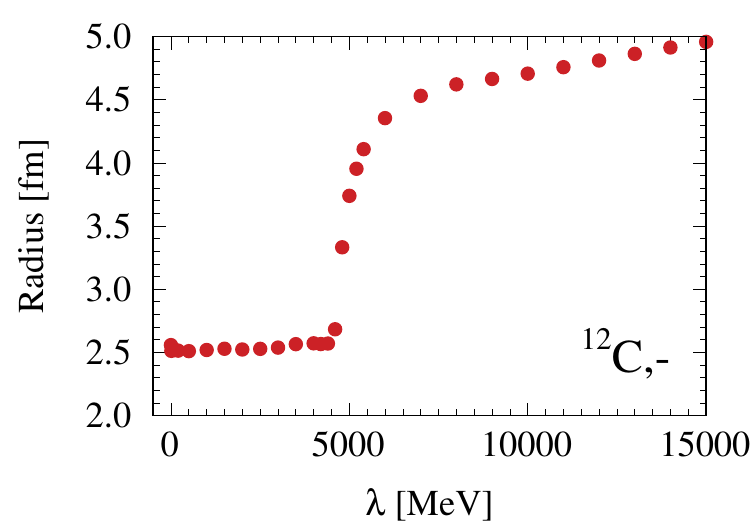}
\caption{
  Intrinsic energy (top) and matter radius (bottom) of $^{12}$C for a negative parity state
  with a basis number of $N_{\rm b}=16$ in the multicool calculation.
  The strength $\lambda$ of the pseudopotential changes.}
\label{fig:ene_negative}
\end{figure}

Figure~\ref{fig:ene_negative} shows the results of the negative parity state of $^{12}$C by increasing $\lambda$ in the pseudopotential.
There are sudden changes in the energy and radius around $\lambda=5000$ MeV, similar to the behavior in the positive parity state,
as shown in Fig.~\ref{fig:ene_positive1}.
Figure~\ref{fig:density3} shows the intrinsic density distributions of the three representative configurations
for the lowest-energy state shown in Table \ref{tab:multi}, and for the excited state with $\lambda=10000$ MeV, respectively.
I confirm the variety of the configurations, some of which exhibit developed $3\alpha$ cluster states.
In the excited state, the configurations show very large radii due to the spatial localization of the $3\alpha$ clusters.

\begin{figure}[t]
\hspace*{-0.00cm}\includegraphics[width=8.60cm,bb=0 0 214 81]{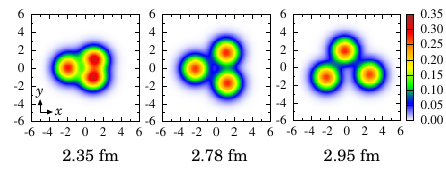}\\ 
\hspace*{-0.60cm}\includegraphics[width=7.95cm,bb=0 0 199 81]{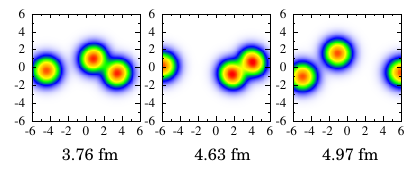}
caption{
  Intrinsic density distributions of the representative configurations of $^{12}$C for negative parity states;
  the lowest-energy state (top three panels) and the excited state with $\lambda=10000$ MeV (bottom three panels).
  Units of the densities with colorbar and the $x$-$y$ axes are in fm$^{-3}$ and in fm, respectively.
  The radius is shown below each panel.
}
\label{fig:density3}
\end{figure}

I obtain the various AMD configurations of $^{12}$C,
characterized by the Gaussian centroid parameter $\bm{Z}$ and the spin amplitudes $\alpha_\uparrow$ and $\alpha_\downarrow$ for each nucleon.
I present these parameters in two representative cases:
the shell-model state with a compact radius of 2.09 fm, as shown in Fig. \ref{fig:density1},
and the $3\alpha$ cluster state with a large radius of 5.03 fm, as shown in Fig. \ref{fig:density2}.
Figure \ref{fig:position} shows the distributions of the real part of $\{\bm{Z}_i\}$ for protons and neutrons in $^{12}$C.
In the shell-model state, all Re($\bm{Z}_i$) values are located near the origin.
In the cluster state, three locations of Re($\bm{Z}_i$) are identified forming an $\alpha$ cluster at each location.

  Figure \ref{fig:spin} shows the absolute squares of the spin amplitudes, $|\alpha_{\uparrow}|^2$ and $|\alpha_{\downarrow}|^2$,
  for each nucleon of $^{12}$C in the two basis states shown in Fig. \ref{fig:position}.
  For each nucleon. $|\alpha_{\uparrow}|^2+|\alpha_{\downarrow}|^2=1$.
  In the cluster state shown in the bottom panel, each nucleon predominantly has either an up or down spin component,
  and four of these nucleons form an $\alpha$ cluster with spin saturation.
  The numbers of nucleons with the up-spin and down-spin are equal to be 6.0 in this basis state.
  For the shell-model state in the top panel, a mixture of the up and down spins is evident in some of nucleons,
  indicating the breaking of the $\alpha$ clusters in $^{12}$C.
  In fact, the numbers of nucleons with the up-spin and down-spin are 2.8 and 9.2, respectively.
  
\begin{figure}[t]
\includegraphics[width=4.10cm,bb=0 0 257 252]{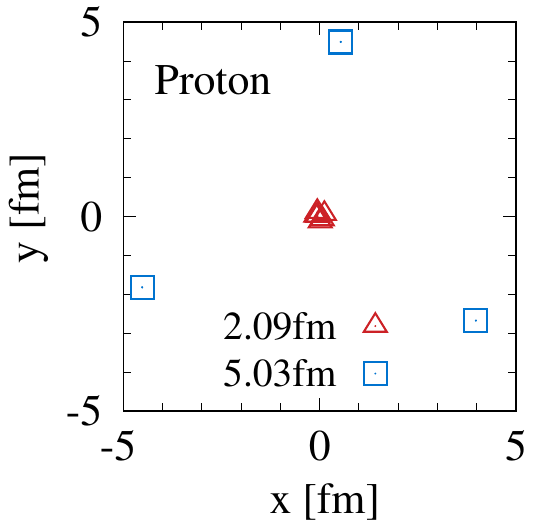}~
\includegraphics[width=4.10cm,bb=0 0 257 252]{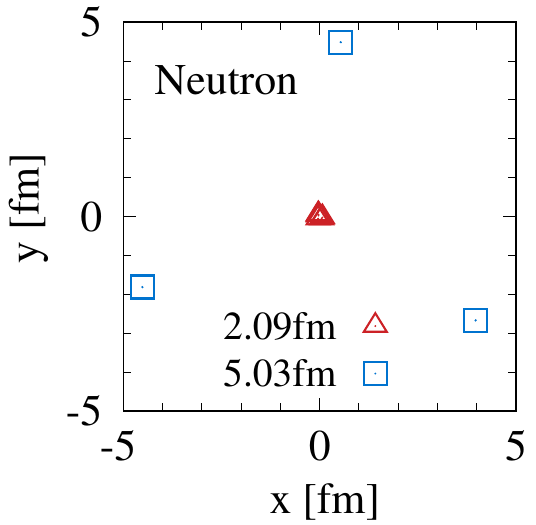}

\caption{
  Real part of the Gaussian centroid parameters $\{\bm{Z}_i\}$ of each nucleon in $^{12}$C
  for the shell-model state (red triangles) with the radius of 2.09 fm shown in Fig. \ref{fig:density1}
  and the cluster state (blue squares) with the radius of 5.03 fm shown in Fig. \ref{fig:density2}.
  Left (right) panel shows protons (neutrons).
}
\label{fig:position}
\end{figure}

\begin{figure}[t]
\includegraphics[width=7.20cm,bb=0 0 362 379]{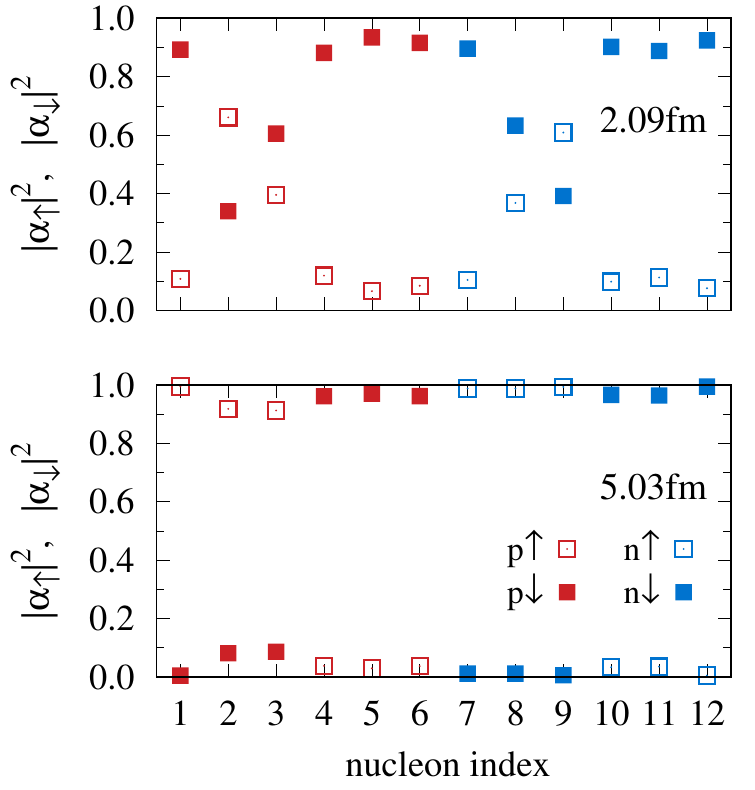}
\caption{
    Absolute squares of the spin amplitudes, $|\alpha_{\uparrow}|^2$ and $|\alpha_{\downarrow}|^2$,
    for each nucleon of $^{12}$C,
    where $|\alpha_{\uparrow}|^2+|\alpha_{\downarrow}|^2=1$.
    The top panel shows the shell-model state with a radius of 2.09 fm, as shown in Fig. \ref{fig:density1}.
    The bottom panel shows the cluster state with a radius of 5.03 fm, as shown in Fig. \ref{fig:density2}.
    The red (blue) squares represent the protons (neutrons).
}
\label{fig:spin}
\end{figure}

\subsection{Energy spectrum}

\begin{table}[t]
  \caption{
    Ground-state properties of $^{12}$C ($0^+_1$); total energy in units of MeV and
    radii of matter ($r_{\rm m}$) and charge ($r_{\rm ch}$)  in units of fm,
    compared with the experimental values \cite{tanihata88,ozawa01,angeli13}.
  }\vspace*{0.1cm}
\label{GS}
\renewcommand{\arraystretch}{1.5}
\begin{tabular}{p{2cm}c p{0.2cm} l p{0.2cm} l }
\hline
           &  Energy    &&  $r_{\rm m}$       && $r_{\rm ch}$ \\
\hline
Experiment &  $-92.16$  &&   2.35(2), 2.31(2) && 2.4702(22)   \\
Present    &  $-88.59$  &&   2.35             && 2.48         \\
\hline
\end{tabular}
\end{table}

I superpose the AMD configurations shown in Figs.~\ref{fig:ene_positive1} ~\ref{fig:ene_positive2} for positive parity
and those in Fig~\ref{fig:ene_negative} for negative parity, with the angular-momentum projection.
I limit the configuration number to approximately 500, which is sufficient for the results to converge.
By solving the eigenvalue problem of the Hamiltonian matrix in Eq.~(\ref{eq:eigen}),
I obtain the ground state of $^{12}$C ($0^+_1$) and list its properties in Table \ref{GS}.
The total energy and the radii of matter and charge are consistent with the experimental values \cite{tanihata88,ozawa01,angeli13}.
The total energy is $-88.6$ MeV, which is slightly underestimated, however, this value is close to those of
other $^{12}$C calculations \cite{fujiwara80,kanada07,imai19,bo20,ichikawa22} that treat the $3\alpha$ cluster states.
The experimental charge radius is reported precisely using the isotope shift, and the present result is close to this value,
where I adopt the charge radii of proton and neutron in the experiments \cite{codata22,atac21}.

\begin{figure}[t]
\centering
\includegraphics[width=8.2cm,bb=0 0 482 657]{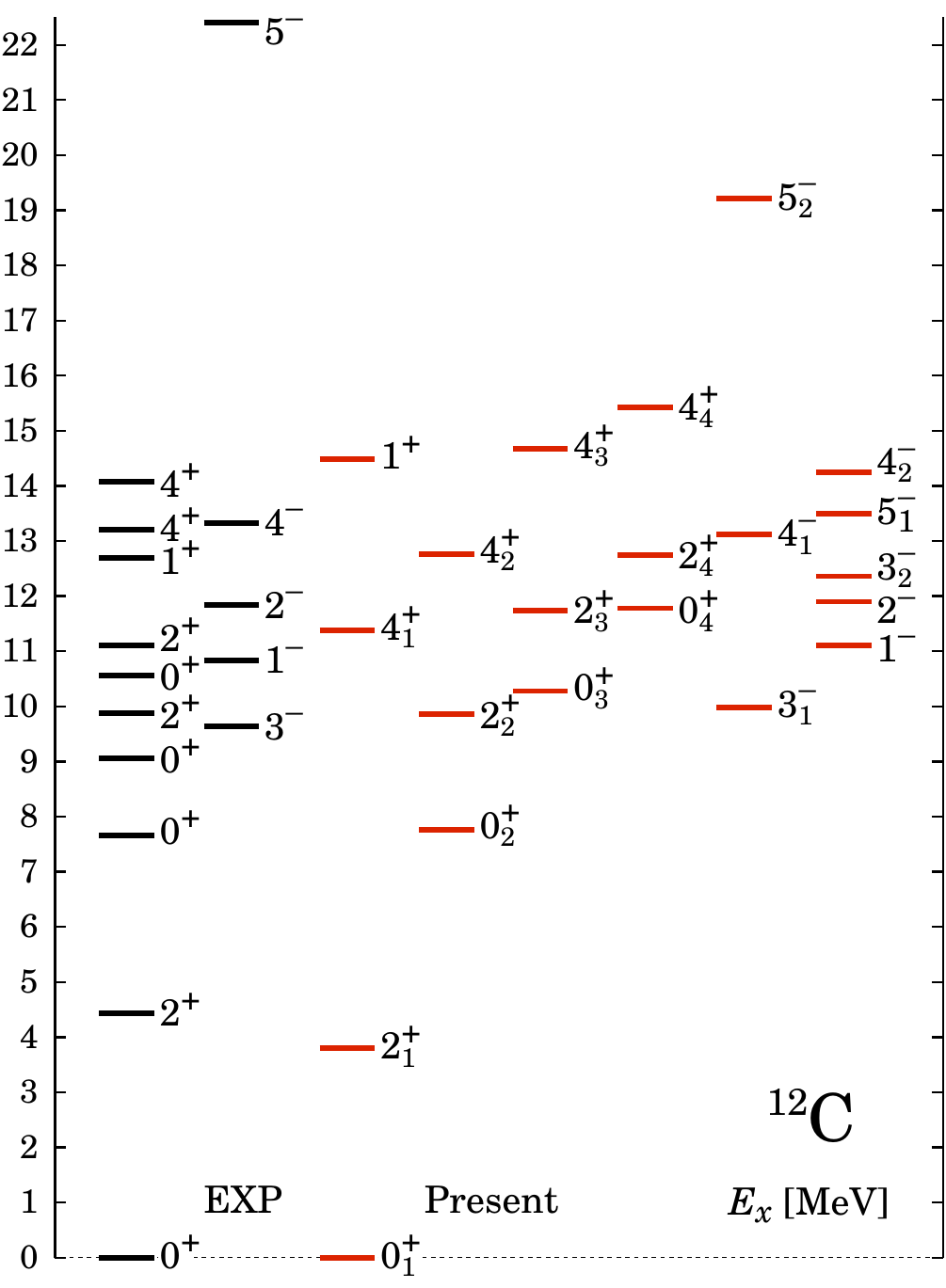} 
\caption{
  Excitation energy spectra of $^{12}$C in the experiments (black lines on the left)
  \cite{ajzenberg90,tunl,kelley17,freer07,hyldegaard10,itoh11,itoh13b,freer11,freer12,marin14,li22a,li22b}
  and the multicool calculation (red lines) in units of MeV.
}
\label{fig:ene_GCM}
\end{figure}

Figure~\ref{fig:ene_GCM} shows the excitation energy spectrum of $^{12}$C,
in comparison with the experimental results \cite{ajzenberg90,tunl,kelley17,itoh11,itoh13b,freer11,freer12,marin14,nndc}.
My results reproduce the experimental levels well.
I confirm the four kinds of the $0^+$ bands with $2^+$ and $4^+$, where the resonances are described within the bound-state approximation.
I identify the individual bands based on the behavior of the configuration mixing, radius, and the transition strengths between the states.
The existence of the four $0^+$ bands is consistent with the conclusions obtained using the $3\alpha$ cluster model
\cite{bo20,funaki15,funaki16,bo16,imai19,takemoto23,kurokawa05,kurokawa07,ichikawa22,cheng25}.
The difference between the present calculation and the $3\alpha$ cluster model is that I do not assume the $\alpha$ clusters in the wave function,
and the breaking of the $\alpha$ clusters is automatically treated in the energy variation.
This breaking yields the contribution of the spin-orbit force in the Hamiltonian for $^{12}$C \cite{itagaki05,suhara13,suhara15}.
As is mentioned, in Refs. \cite{kanada07,chernykh07,fukuoka13,suhara15}, their results show three kinds of the $0^+$ bands, whereas I confirm four.
I attribute this difference to the fact that, in the multicool method, we generate the AMD configurations with very large radii,
using the orthogonal condition with the pseudopotential, as shown in Figs. \ref{fig:ene_positive1} and \ref{fig:ene_positive2}.
  This aspect is important for describing the spatially extended states, such as the fourth bands,
  as is explained in detail later using Table \ref{tab:quanta}.

  I mention the limitations of the present model despite the reproduction of the energy spectrum.
  One is the treatment of unbound states to distinguish the resonances embedded in the continuum states,
  since cluster states often exist above the threshold energies of the cluster emissions.
  At present I use the bound-state approximation, but further development of the framework would be desirable,
  such as combining the method with the complex scaling \cite{myo14,zhang22,myo23a,kurokawa07}.
  Secondly, the present model does not take into account the effects of the tensor force and
  short-range repulsion in the nucleon-nucleon interaction;
  instead we use an effective interaction that is softened without high-momentum components.
  The explicit effect of the tensor force is of particular interest.
  So far, I have developed the tensor-optimized antisymmetrized molecular dynamics \cite{myo17,lyu20,myo22},
  in which realistic nuclear interactions can be treated in terms of correlation functions multiplied to the AMD wave function,
  although this method would increase the numerical effort.

\begin{table}[t]
\centering
\caption{
  Expectation values of the principal quantum number operator $N$ of the states of $^{12}$C
  and their matter radii in units of fm in the multicool calculation.
  I also present the radii of the $3\alpha$ model with THSR \cite{funaki15},
  in which the $0^+_{3,4}$ states are exchanged.
}\vspace*{0.1cm}
\label{tab:quanta}
\renewcommand{\arraystretch}{1.5}
\begin{tabular}{lcrrrrrrrrrrrrrrrrrrrr}
\hline
                    && $0^+_1$ && $0^+_2$ && $0^+_3$ && $0^+_4$  \\
\hline
$\bra N\ket$        && 9.55    && 28.60   && 39.17   && 47.63    \\
Radius              && 2.35    &&  3.63   &&  4.10   &&  4.48    \\
THSR                && 2.4     &&  3.7    &&  4.2    &&  4.7     \\
\hline
\end{tabular}
\hspace*{1.0cm}
\\[0.1cm]
\begin{tabular}{lcrrrrrrrrrrrrrrrrrrrr}
\hline
                    && $2^+_1$  && $2^+_2$ && $2^+_3$ && $2^+_4$  \\
\hline
$\bra N\ket$        && 10.02    && 36.66   && 36.71   &&  50.83   \\
Radius              &&  2.41    && 4.02    &&  4.02   &&   4.59   \\
THSR                &&  2.4     && 3.9     &&         &&          \\
\hline
\end{tabular}
\hspace*{1.0cm}
\\[0.1cm]
\begin{tabular}{lcrrrrrrrrrrrrrrrrrrrr}
\hline
                     && $4^+_1$ && $4^+_2$ && $4^+_3$ && $4^+_4$ && $1^+$     \\
\hline
$\bra N\ket$         && 11.69   &&  41.87  && 52.76   && 77.20   && 10.05     \\
Radius               &&  2.55   &&   4.24  &&  4.67   &&  5.53   &&  2.43     \\
THSR                 &&  2.3    &&   4.5   &&         &&         &&           \\
\hline
\end{tabular}
\\[0.1cm]
\begin{tabular}{lcrrrrrrrrrrrrrrrrrrrr}
\hline
              & $1^-$   & $2^-$  & $3^-_1$ & $3^-_2$  & $4^-_1$ & $4^-_2$  & $5^-_1$ & $5^-_2$  \\
\hline                                                                          
$\bra N\ket$  & 43.11   & 45.94  & 20.09   & 74.43    & 29.96   & 65.34    & 79.86   & 39.37    \\
Radius        &  4.29   &  4.41  &  3.17   &  5.44    &  3.69   &  5.13    &  5.62   &  4.10    \\
\hline
\end{tabular}
\end{table}

I discuss the detailed properties of the energy levels.
Table~\ref{tab:quanta} lists the expectation values of the principal quantum number operator $N$ of the harmonic oscillator (HO).
The operator $N$ is defined in terms of the kinetic energy and radius,
and $\hbar \omega=2\nu \hbar^2/m$ where $m$ is the nucleon mass \cite{kanada07}:
\begin{equation}
  \begin{split}
  N\coloneqq \sum_{i=1}^A \left(\dfrac{\bm{p}_i^2}{4\hbar^2\nu}+\nu \bm{r}^2_i - \dfrac32 \right).
  \end{split}
  \label{eq:quanta}
\end{equation}
The values of $\bra N \ket$ are useful for estimating the amounts of the excitation in the picture of the harmonic-oscillator shell model.
I also present the matter radii of the states to evaluate their spatial extension and the development of the clusters.

In $^{12}$C, the configuration of the lowest HO quanta of $\bra N\ket=8$ yields a radius of $\sqrt{49/(48\nu)}$, which leads to 2.08 fm for reference.
In the present ground state, the matter radius is 2.35 fm, which is 0.27 fm larger than the lowest value.
This indicates the inclusion of the spatial correlations of deformation or clustering beyond the $p$ shell in the ground state,
as is discussed later based on the density distributions shown in Fig. \ref{fig:density5}.
In Table~\ref{tab:quanta}, it is found that the first band consisting of $0^+_1$, $2^+_1$, and $4^+_1$
have the compact radii of less than 3 fm. This fact means a shell-model state with lower quanta.
The $1^+$ state also exhibits a similar result.
The other states show the large radii beyond 3 fm and the large HO quanta.
The $0^+_2$ Hoyle state shows the large radius of 3.63 fm and the large quanta of 28.6 in the present calculation.
The $0^+_{3,4}$ states show very large radii beyond 4 fm.
The $0^+_4$ band shows the largest radii and quanta among the same spin states.
I also compare the radii of the present calculation with those obtained
in the $3\alpha$ condensate THSR (Tohsaki-Horiuchi-Schuck-R\"opke) model \cite{funaki15}.
My radii are similar to their values in the individual states.
It is noted that throughout this paper, the order of the $0^+_3$ and $0^+_4$ bands in
the $3\alpha$ cluster calculations \cite{bo16,funaki15,funaki16,imai19,ichikawa22,takemoto23,cheng25} is reversed
for the correspondence of the states obtained in the present calculation. 
This is related to the assignment of the breathing mode and also the effect of the spin-orbit force.
A detailed explanation will be provided later.

\begin{figure}[t]
\hspace*{-0.00cm}\includegraphics[width=8.60cm,bb=0 0 213 79]{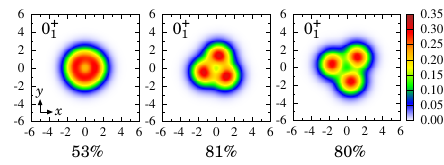}\\ 
\hspace*{-0.65cm}\includegraphics[width=7.95cm,bb=0 0 199 80]{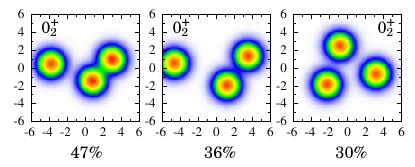}\\
\hspace*{-0.65cm}\includegraphics[width=7.95cm,bb=0 0 199 80]{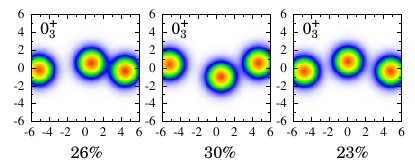}\\
\hspace*{-0.65cm}\includegraphics[width=7.95cm,bb=0 0 199 80]{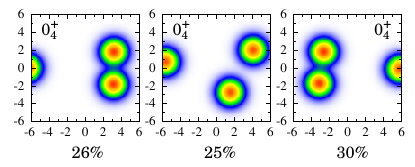}
\caption{
  Intrinsic density distributions of the representative configurations in the $0^+_{1-4}$ states of $^{12}$C.
  Units of densities with colorbar and the $x$-$y$ axes are in fm$^{-3}$ and in fm, respectively.
  The values below each panel represent the weights in the total wave functions, expressed as a percentage.
}
\label{fig:density5}
\end{figure}

I discuss the internal structures of the four $0^+$ bands in terms of their densities.
Figure~\ref{fig:density5} shows the intrinsic density distributions of the three representative configurations
for each of the $0^+_{1-4}$ states of $^{12}$C. The weights of the configurations in the total wave functions are also shown below each panel.
In the ground $0^+_1$ state, the spatially compact shell-model configuration and the compact $3\alpha$ cluster configuration
are dominant with large weights and they are mixed.
In the $0^+_2$ Hoyle state, the $3\alpha$ configuration develops and the $^8$Be-$\alpha$ structure is confirmed,
where the $\alpha$-$\alpha$ distance of $^8$Be is changeable and can be a generator coordinate.
In the $0^+_3$ state, the $3\alpha$ clusters form a linear-chain with a slight bending,
which is a common feature in other studies \cite{kurokawa07,kanada07,suhara15,bo16,imai19,ichikawa22,takemoto23,cheng25}.
It is noted that this state is assigned as the $0^+_4$ state in the $3\alpha$ cluster calculations.
In the present $0^+_4$ state, the $^8$Be-$\alpha$ structure is spatially more significant with a large relative distance
between them \cite{bo16,imai19,ichikawa22,takemoto23,cheng25}.
In addition, the $^8$Be part is also extendable in the $\alpha$-$\alpha$ distance.
This feature of the $0^+_4$ state can be interpreted as the breathing mode of the $0^+_2$ state,
which is predicted by the past $3\alpha$ cluster model as the $0^+_3$ state \cite{bo16}.
In fact, the present $0^+_4$ state has the largest radius, as shown in \ref{tab:quanta},
which is consistent with the previous results of the $3\alpha$ cluster model \cite{bo16,funaki15,funaki16,imai19,ichikawa22,takemoto23}.

It is noted that the weights of the main configurations in the $0^+_{2-4}$ states are less than 50\% and are not very large.
This is because these states are spatially extended with large radii exceeding 3 fm, as is shown in Table~\ref{tab:quanta}.
This property indicates that a single AMD configuration consisting of the Gaussian wave packets of nucleons is insufficient
to represent a spatially extended state, and the superposition of the configurations is necessary.
In the multicool calculation, the AMD configurations are optimized through superposition.
This superposition plays an important role in expressing the spatial extension of the relative distances between the $\alpha$ clusters
in the $0^+_{2-4}$ states. 
In particular, the relative wave function of the $\alpha$-$\alpha$ system is a Gaussian wave packet in AMD,
and the superposition of the AMD configurations optimizes the relative wave function between clusters.
This optimization process is performed smoothly in the multicool variation
to create the spatially extended relative wave functions in the $0^+_{2-4}$ states.

\begin{table}[t]
\centering
\caption{
  Contributions of the spin-orbit (LS) force $\bra \widetilde V_{\rm LS}\ket$ in $^{12}$C
  normalized by the ground-state ($0^+_1$) value of $-11.96$ MeV,
  and the expectation values of the squared spin operator $\bm{S}^2$, where $\bm{S}$ is a sum of the nucleon spin operator.
  }\vspace*{0.1cm}
\label{spin}
\renewcommand{\arraystretch}{1.5}
\begin{tabular}{lcrrrrrrrrrrrrrrrrrrrr}
\hline
                                 && $0^+_1$ && $0^+_2$ && $0^+_3$ && $0^+_4$ \\
\hline                                     
$\bra \widetilde V_{\rm LS}\ket$ && 1.00    && 0.25    && 0.36    && 0.03    \\
$\bra \bm{S}^2 \ket$             && 0.46    && 0.16    && 0.26    && 0.01    \\
\hline
\end{tabular}
\\[0.1cm]
\begin{tabular}{lcrrrrrrrrrrrrrrrrrrrr}
\hline
                                 && $2^+_1$ && $2^+_2$ && $2^+_3$ && $2^+_4$  \\
\hline                 
$\bra \widetilde V_{\rm LS}\ket$ && 0.63    && 0.06    && 0.10    &&  0.07    \\
$\bra \bm{S}^2 \ket$             && 0.27    && 0.03    && 0.05    &&  0.03    \\
\hline
\end{tabular}
\\[0.1cm]
\begin{tabular}{lcrrrrrrrrrrrrrrrrrrrr}
\hline
                                 && $4^+_1$ && $4^+_2$ && $4^+_3$ && $4^+_4$  \\
\hline                              
$\bra \widetilde V_{\rm LS}\ket$ &&  0.37   &&  0.05   &&  0.02   && 0.03     \\
$\bra \bm{S}^2 \ket$             &&  0.14   &&  0.02   &&  0.01   && 0.02     \\
\hline
\end{tabular}
\end{table}

\subsection{$\alpha$-cluster breaking}

In Ref. \cite{suhara15}, the authors superpose the $3\alpha$ cluster basis states and the $p_{3/2}$ sub-closed configuration
as the $\alpha$-cluster breaking.
They discuss how the spin-orbit force affects the level order of the $0^+$ states of $^{12}$C
due to the coupling with the $p_{3/2}$ sub-closed configuration. 
In the $3\alpha$ cluster model without the spin-orbit force, the $0^+_4$ state, namely, breathing mode of the $0^+_2$ state,
has a lower energy than that of the $0^+_3$ state, which has a linear-chain structure \cite{bo16}.
In relation to the present results, treating the spin-orbit force as an $\alpha$-cluster breaking can change
the level order of the $0^+_3$ and $0^+_4$ states.

I discuss the $\alpha$-cluster breaking in the $0^+_{1-4}$ band states of $^{12}$C.
The spin-orbit force in the Hamiltonian becomes a key for this purpose.
If the $\alpha$ cluster forms with an $s$-wave closed configuration, its spin-orbit contribution is zero in $^{12}$C.
I also calculate the expectation values of the squared spin operator $\bm{S}^2$,
where $\bm{S}=\sum_{i=1}^A \bm{s}_i$ with nucleon spin operator $\bm{s}_i$ \cite{kanada07}.
The values of $\bra \bm{S}^2 \ket$ are useful for discussing the spin component of the state.
If the $\alpha$ cluster forms in a nucleus, its contribution to $\bra \bm{S}^2 \ket$ becomes zero, similar to the spin-orbit force.
Hence, I focus on the two quantities associated with the spin to estimate the $\alpha$-cluster breaking.

Table~\ref{spin} shows the contributions of the spin-orbit force to the total energy,
as well as $\bra \bm{S}^2 \ket$ for each $^{12}$C state.
The spin-orbit (LS) contributions $\bra \widetilde V_{\rm LS}\ket$ are normalized by the ground-state ($0^+_1$) value of $-11.96$ MeV as a criterion.
This value is the largest attraction among the $^{12}$C states due to the $p_{3/2}$ sub-closed nature.
It is found that the $0^+_{2,3}$ states give $\bra \widetilde V_{\rm LS}\ket$ of 0.25 and 0.36, respectively,
representing the attractions of $-3.0$ MeV and $-4.3$ MeV.
These finite values suggest the degree to which the $\alpha$-cluster breaks down in the two states.
On the other hand, the $0^+_4$ state gives a tiny value of 0.03, indicating a rather pure $3\alpha$ cluster state as confirmed in the densities.
The values of $\bra \bm{S}^2\ket$ for the $0^+_{1-4}$ states follow the results of $\bra \widetilde V_{\rm LS}\ket$;
the value of $0^+_1$ is 0.46, which could serve as a criterion, and the $0^+_{2,3}$ give the finite values.
The $0^+_4$ gives a tiny value of 0.01 close to zero.

I consider the reason for the notable values of $\bra \widetilde V_{\rm LS}\ket$ in the $0^+_{2,3}$ states. 
In these two states, there are mixings of the shell-model states,
such as the spherical configuration of the top-left panel of Fig.~\ref{fig:density5} for the $0^+_1$ state.
This configuration gives the excitation energy of 10.2 MeV,
which is close to the energies of the $0^+_{2,3}$ states as shown in Fig. \ref{fig:ene_GCM}.
It also yields the HO quanta $\bra N\ket$ of 8.0 being the lowest value, and $\bra \widetilde V_{\rm LS}\ket$ of 2.4
and $\bra \bm{S}^2\ket$ of 1.6, representing the significant effect of spin.
The weights of this configuration are 9\% and 11\% for the $0^+_{2}$ and $0^+_{3}$ states, respectively, but only 0.5\% for the $0^+_4$ state.
The differences between the three states can explain the behavior of the spin quantities shown in Table~\ref{spin}.
A similar discussion is given in Ref.~\cite{suhara15}.

For the $2^+$ states, the $2^+_1$ state exhibits $\bra \widetilde V_{\rm LS}\ket$ of 0.63,
regarding the shell-model state, and the remaining $2^+_{2-4}$ states have very small values
as do the $\bra \bm{S}^2\ket$ values.
Based on these results, $3\alpha$ cluster formation is prevalent in the $2^+_{2-4}$ states.
A similar property to that of the $2^+$ states is observed in the $4^+$ states.
The $4^+_1$ state has a relatively large contribution of the spin-orbit force, though it is rather weak compared to the $2^+_1$ state.
The other $4^+_{2-4}$ states provide the minor contributions.
The $\bra \bm{S}^2\ket$ values follow this trend; the $4^+_{2-4}$ states provide small values indicating a pure $3\alpha$ cluster state.

From the analysis in terms of $\bra \widetilde V_{\rm LS}\ket$ and $\bra \bm{S}^2\ket$,
I conclude that the pure $3\alpha$ cluster structure is preferable in the $0^+_4$, $2^+_{2-4}$, and $4^+_{2-4}$ states.
The first band consisting of $0^+_1$, $2^+_1$, and $4^+_1$ involves the shell-model states that result in the large contributions of the spin-orbit force.
For the $0^+_{2,3}$ states, I observe some amounts of the $\alpha$-cluster breaking,
which induce the spin-orbit attractions of a few MeV in the total energy.
This attraction can explain why the $0^+_3$ linear-chain state has a lower excitation energy than the $0^+_4$ breathing excited state
in the present calculation, differently from the $3\alpha$ cluster calculations without the spin-orbit effect.

I further estimate the $\alpha$-cluster breaking in the $0^+_{1-4}$ band states
by considering their overlap with the $3\alpha$ Brink-Bloch (BB) wave functions \cite{brink66}.
For this purpose I construct an orthonormalized basis set spanning solely the $3\alpha$ space.
The $3\alpha$ BB basis states are given as follows:
\begin{eqnarray}
  \begin{split}
    \Phi_{\rm BB} &= P^{J}_{MK}P^+ {\cal A}\{ \varphi_\alpha(\bm{R}_1),\varphi_\alpha(\bm{R}_2),\varphi_\alpha(\bm{R}_3) \},
    \\
    \varphi_\alpha(\bm{R}) &= \phi_{\bm{R}}\, \phi_{\bm{R}}\, \phi_{\bm{R}}\, \phi_{\bm{R}} \;
    \chi_{p\uparrow} \chi_{p\downarrow} \chi_{n\uparrow} \chi_{n\downarrow},
      \\
    \phi_{\bm{R}} &= \left(\frac{2\nu}{\pi}\right)^{3/4} e^{-\nu(\bm{r}-\bm{R})^2},
  \end{split}
\end{eqnarray}
where the same value of $\nu$ in the Gaussian is adopted as used in Eq.~(\ref{eq:AMD}).
I simply write the BB basis states omitting the spin-parity notation of $J^+$.
The real parameters $\bm{R}_{1,2,3}$ determine the positions of the $3\alpha$ clusters.
According to Ref. \cite{suhara15}, the $3\alpha$ BB basis states are prepared using the intercluster distances $d_1$ and $d_2$
with $d_1 \le d_2$ and the bending angle $\theta$, as shown in Fig. \ref{fig:BB}.
I set $d_{1,2}=1,2,\ldots,12$ fm and $\theta= n\pi/12$ with $n=1,2,\ldots,12$ to converge the results.

By diagonalizing the overlap matrix of the BB basis states for each $J^+$ state,
I construct the orthonormalized $3\alpha$ basis states $\{\Phi^{3\alpha}_p\}$ with the index $p=1,\ldots,N_{3\alpha}$,
by superposing the BB basis states with the index $i$:
\begin{eqnarray}
  \begin{split}
    \Phi^{3\alpha}_p &= \sum_i c^p_i\, \Phi_{{\rm BB},i},\quad
    \langle \Phi^{3\alpha}_p | \Phi^{3\alpha}_q \rangle = \delta_{pq}.
  \end{split}
\end{eqnarray}
The $3\alpha$ components in the $0^+_{1-4}$ band states are obtained
as the squared overlap, $\sum_p^{N_{3\alpha}} \bigl|\langle \Phi^{3\alpha}_p| \Psi_{\rm t}^{J^+} \rangle\bigr|^2$.

The results are shown in Table \ref{tab:BB}.
The $0^+_1$ ground state exhibits the largest component of the $\alpha$-cluster breaking of 23\%.
whereas the $0^+_{2,3}$ states exhibit the breaking of around 10\%, and the $0^+_4$ state exhibits a negligible breaking.
These values are consistent with the results of $\bra \widetilde V_{\rm LS}\ket$ and $\bra \bm{S}^2\ket$ shown in Table \ref{spin}.
The $2^+_1$ and $4^+_1$ states exhibit the breaking components that are smaller than the $0^+_1$ case.
The $2^+_{2-4}$ and $4^+_{2-4}$ states exhibit the very small components of breaking with a few percent,
which support the $3\alpha$ cluster picture.

\begin{figure}[t]
\includegraphics[width=5.0cm,bb=0 0 245 99]{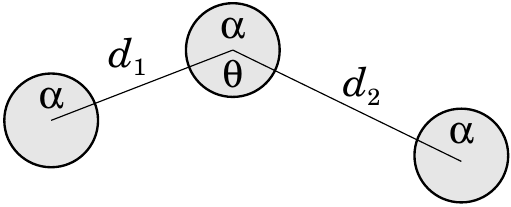}
\caption{
  Schematic configuration of $3\alpha$ clusters for $^{12}$C with the intercluster distances $d_1$ and $d_2$ and the bending angle $\theta$.
}
\label{fig:BB}
\end{figure}
%

\begin{table}[t]
  \caption{
    Squared overlap of the $0^+_{1-4}$ band states with the $3\alpha$ wave functions.
  }\vspace*{0.1cm}
\label{tab:BB}
\renewcommand{\arraystretch}{1.5}
\begin{tabular}{p{0.8cm} p{0.8cm} p{0.8cm} p{0.8cm} p{0.8cm} }
\hline
        & $J^+_1$& $J^+_2$   &  $J^+_3$   & $J^+_4$ \\
\hline
$0^+$   & 0.77   &   0.91    &   0.89     &  0.99   \\
$2^+$   & 0.85   &   0.98    &   0.97     &  0.98   \\
$4^+$   & 0.92   &   0.98    &   0.95     &  0.97   \\
\hline
\end{tabular}
\end{table}

\subsection{Charge form factor}

\begin{figure}[t]
\includegraphics[width=8.2cm,bb=0 0 360 252]{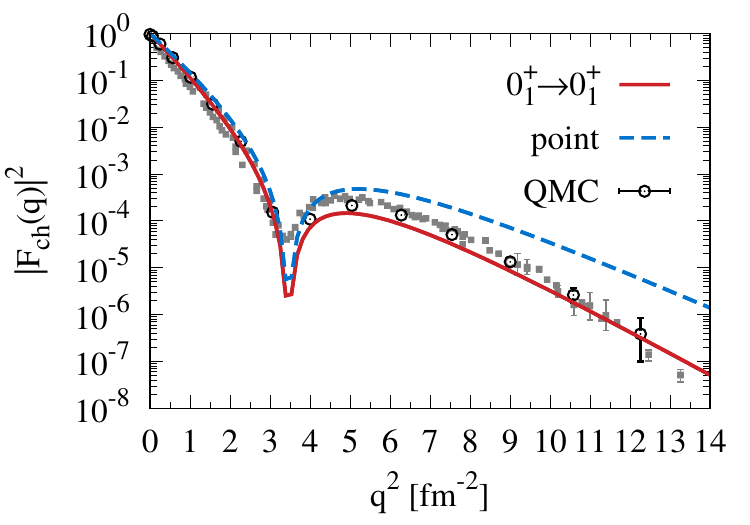}\\[0.1cm]
\includegraphics[width=8.2cm,bb=0 0 360 252]{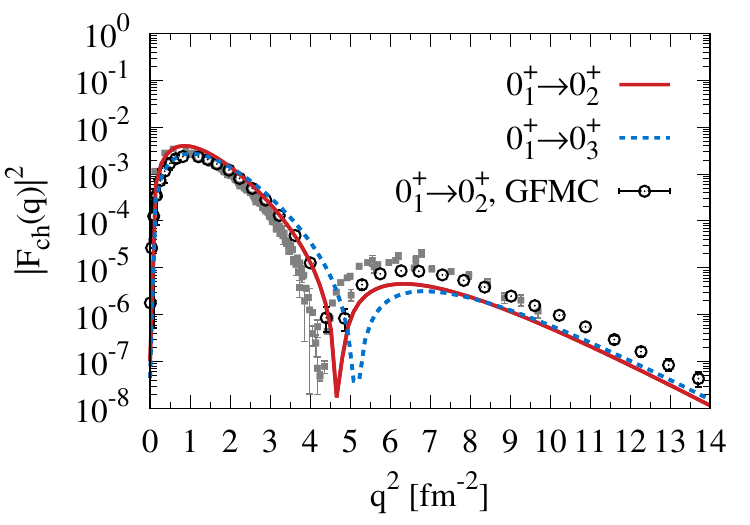}
\caption{
  Absolute square of the charge form factors $|F_{\rm ch}(q)|^2$ of $^{12}$C
  in the elastic ($0^+_1 \to 0^+_1$, top) and the inelastic ($0^+_1 \to 0^+_2$, bottom) channels
  with solid lines in comparison with the experiments (squares) \cite{strehl68,strehl70,sick70,nakada71a,nakada71b,chernykh10}.
  The dashed line in the top panel represents the point-proton distribution $|F_{\rm p}(q)|^2$.
  The open circles represent the results of the Monte Carlo (QMC and GFMC) calculations \cite{lovato13,carlson15}.
  The bottom panel also shows the inelastic $0^+_1 \to 0^+_3$ channel with a dotted line.
}
\label{fig:form}
\end{figure}
I discuss the reliability of the present $0^+$ states of $^{12}$C using the charge form factors $F_{\rm ch}(q)$
with the transfer momentum $q$, as observed in the electron scattering.
I evaluate the absolute square of $F_{\rm ch}(q)$ using the point-proton distribution $F_{\rm p}(q)$ and
the proton size effect \cite{tassie58,uegaki79,kamimura81,funaki06} as:
\begin{eqnarray}
  \begin{split}
    |F_{\rm ch}(q)|^2 &=|F_{\rm p}(q)|^2 \cdot e^{-\langle r^2_p \rangle q^2/3},
  \end{split}
\end{eqnarray}
where the proton radius $\sqrt{\langle r^2_p \rangle}$ is set to 0.841 fm \cite{codata22},
as used in the charge radius in Table~\ref{GS}.

Figure~\ref{fig:form} shows $|F_{\rm ch}(q)|^2$ for the elastic ($0^+_1 \to 0^+_1$) and
the inelastic ($0^+_1 \to 0^+_2$ and $0^+_1 \to 0^+_3$) channels.
I also present the point-proton distribution $|F_{\rm p}(q)|^2$ for the elastic case.
I find that the two distributions of $|F_{\rm ch}(q)|^2$ for $0^+_1 \to 0^+_1$ and $0^+_1 \to 0^+_2$ are consistent with the experimental data.
A more detailed comparison reveals that, for the elastic case, the present distribution reproduces the experiments in the overall momentum region.
This is related to the reproduction of the charge radius of the $0^+_1$ state as shown in Table~\ref{GS}.
In the distribution, a slight underestimation compared to the experiments is observed in the higher-momentum region beyond $q^2=4$ fm$^{-2}$,
which is discussed later.
For comparison, the quantum Monte Carlo (QMC) calculation reproduces the experimental distribution including high-momentum region \cite{lovato13}.

For the $0^+_1 \to 0^+_2$ inelastic channel, the dip position obtained in the present calculation is 4.7 fm$^{-2}$, 
which is slightly higher than the experimental value of 4.2 fm$^{-2}$.
In the higher momentum region beyond the dip, my results give the underestimation
and this tendency is commonly observed in the calculations of algebraic cluster model \cite{bijker20},
AMD \cite{kanada07,kanada09}, Fermionic molecular dynamics \cite{chernykh07}, and density functional theory \cite{fukuoka13}.
The Green's function Monte Carlo (GFMC) calculation reproduces the experimental distribution including high momentum region \cite{carlson15}.
It is remarked that the $3\alpha$ condensate THSR calculations reproduce the two distributions well \cite{funaki06,chernykh07,bo20},
although it does not take into account the $\alpha$-cluster breaking and the effect of the spin-orbit force.

Regarding the underestimation of my results in the high-momentum region of the two distributions,
the tensor correlation induced by realistic nuclear forces has a possibility to increase the high-momentum component
of around $q \simeq 2-3$ fm$^{-1}$ \cite{ogawa06,lyu20,myo22}.
In this sense, the common underestimation of the present distributions in the high-momentum region might be meaningful.
It would be interesting to examine the effect of tensor correlation on the form factors of $^{12}$C \cite{carlson15,ogawa06}.

For the inelastic channel of $0^+_1\to 0^+_3$, the linear-chain state,
the dip position shifts to a higher momentum direction compared to the $0^+_2$ distribution, by about 0.5 fm$^{-2}$.
This tendency is commonly observed in other calculations \cite{uegaki79,kanada07}, but, there is no experimental data with which to compare it.
\subsection{Monopole transition}
I discuss the monopole transitions of $^{12}$C, which are in general suggested to be enhanced by the cluster development in a nucleus \cite{yamada08}.
In the previous $3\alpha$ cluster calculations \cite{bo16,cheng25},
the authors investigated the monopole transitions between four kinds of the $0^+$ band states.
In the present analysis, I reverse the third and fourth $0^+$ bands assigned in their papers,
according to the internal structures of the two bands obtained in the present calculation.
In Ref. \cite{bo16}, the authors confirm the strong transitions of $0^+_2 \to 0^+_4$ in $^{12}$C.
This tendency is also evident in the $2^+_2 \to 2^+_4$, and $4^+_2 \to 4^+_4$ transitions \cite{cheng25}.
They conclude that the fourth band, consisting of $0^+_4$, $2^+_4$, and $4^+_4$, is interpreted as the breathing mode of
the second band, consisting of $0^+_2$, $2^+_2$, and $4^+_2$.
It would be interesting to investigate this property in the present calculation,
in which the $\alpha$ cluster is not assumed and the spin-orbit force can contribute to the states.

\begin{table}[t]
  \caption{
    Magnitudes of the matrix elements of the electric monopole transitions $|M(E0)|$ of $^{12}$C in units of $e\,$fm$^2$.
    I compare the results with those of $3\alpha$ models with THSR \cite{bo16} and CNN \cite{cheng25} as well as
    $3\alpha$+$p_{3/2}$ \cite{suhara15} and the experimental data \cite{ajzenberg90,tunl,kelley17}.
  }\vspace*{0.1cm}
\label{tab:mono}
\renewcommand{\arraystretch}{1.5}
\begin{tabular}{lcccccccccccc}
\hline
                  & Present    &  THSR   & CNN       & $3\alpha$+$p_{3/2}$ & Expt \\
\hline                                             
$0^+_1 \to 0^+_2$ &  6.93      &  6.24   &  6.08     & 8.1                 & 5.48(22)\\
$0^+_1 \to 0^+_3$ &  4.74      &  3.64   &  3.48     & 2.1                 &       \\
$0^+_1 \to 0^+_4$ &  0.25      &  3.60   &  3.25     &                     &       \\
$0^+_2 \to 0^+_3$ & 28.90      &  1.21   &  2.11     & 16.1                &       \\
$0^+_2 \to 0^+_4$ & 12.31      & 46.97   & 42.89     &                     &       \\
$0^+_3 \to 0^+_4$ &  9.38      &  7.69   & 20.78     &                     &       \\
\hline
\end{tabular}
\\[0.1cm]
\begin{tabular}{lccc|ccccccccc}
\hline
                  & Present  &  CNN       &&                   & Present  &  CNN       \\
\hline                                               
$2^+_1 \to 2^+_2$ &  5.11    &  5.51      && $4^+_1 \to 4^+_2$ & 13.89    &  2.24      \\
$2^+_1 \to 2^+_3$ &  5.78    &  0.46      && $4^+_1 \to 4^+_3$ &  5.05    &  2.97      \\
$2^+_1 \to 2^+_4$ &  2.62    &  4.21      && $4^+_1 \to 4^+_4$ &  0.59    &  3.33      \\
$2^+_2 \to 2^+_3$ &  6.50    &  2.67      && $4^+_2 \to 4^+_3$ & 13.40    & 25.34      \\
$2^+_2 \to 2^+_4$ & 31.90    & 57.84      && $4^+_2 \to 4^+_4$ & 32.04    & 56.45      \\
$2^+_3 \to 2^+_4$ &  3.91    &  1.30      && $4^+_3 \to 4^+_4$ & 14.35    & 51.43      \\
\hline
\end{tabular}
\end{table}

Table~\ref{tab:mono} shows the magnitudes of the matrix elements of the electric monopole transitions, $|M(E0)|$, of $^{12}$C
for four kinds of the $0^+$ band states.
I compare my results with the calculations of the $3\alpha$ cluster models using the $\alpha$-condensate THSR \cite{bo16} and the
control neutral network (CNN) \cite{cheng25}, as well as  the $3\alpha$ cluster model adding the $p_{3/2}$-closed configuration \cite{suhara15},
which is denoted as $3\alpha+p_{3/2}$.
For the $0^+$ transitions, the present value of $0^+_1 \to 0^+_2$ is 6.93 $e$ fm$^2$,
which is consistent with other calculations and the experimental data of 5.48(22) $e$ fm$^2$ \cite{ajzenberg90,tunl,kelley17}.

In the current calculation of the $0^+$ states, the $0^+_2 \to 0^+_3$ transition has the largest value
and the $0^+_2 \to 0^+_4$ transition is the second largest.
In Refs. \cite{bo16,cheng25}, the $0^+_2 \to 0^+_4$ transition is the largest value and the $0^+_2 \to 0^+_3$ value is small.
Other $3\alpha$ model calculations commonly exhibit this tendency \cite{funaki15,takemoto23}.
In Ref. \cite{suhara15}, the $\alpha$-cluster breaking is introduced and
the authors obtain a large value of $0^+_2 \to 0^+_3$, which is the same feature as my results.
In my results, the cluster breaking due to the spin-orbit force affects
the $0^+_2$ state, but not the $0^+_4$ state, as shown in Table~\ref{spin}, and
the different breakings between the two states is considered to reduce the $0^+_2 \to 0^+_4$ transition, 
although the value of 12.3 $e$fm$^2$ is still significant.
This could be a possible signature of the breathing mode of the $0^+_4$ state excited from the $0^+_2$ state,
though it is not significantly evident in the transition strength compared to the results of the $3\alpha$ cluster model.
On the other hand, the $0^+_2 \to 0^+_3$ transition is enhanced in my results and also in the $3\alpha+p_{3/2}$ calculation \cite{suhara15}.
This is because the $0^+_{2,3}$ states commonly contain the $\alpha$-cluster breaking in the present results.
The $0^+_3 \to 0^+_4$ transition is not enhanced as much, which differs from the $3\alpha$ CNN results \cite{cheng25},
but agrees with those in THSR \cite{bo16}.
This difference is related to the different effects of the $\alpha$-cluster breaking in the two states;
the $0^+_3$ state is finite, but the $0^+_4$ state is negligible.

For the $2^+$ and $4^+$ states, the monopole transitions of $2^+_2 \to 2^+_4$ and $4^+_2 \to 4^+_4$ are
the largest among the $2^+$ and $4^+$ states, respectively in the present calculations.
These characteristics are the same as those obtained in the $3\alpha$ calculation \cite{cheng25}.
We can recognize the breathing excitation of the $2^+_4$ and $4^+_4$ states from the $2^+_2$ and $4^+_2$ states, respectively.
In the present calculation, the $\alpha$cluster breaking is minimal in the $2^+_{2-4}$ and $4^+_{2-4}$ states compared to the $0^+_{2,3}$ states.
Therefore, it is reasonable that the present results agree with those in $3\alpha$ CNN \cite{cheng25}.

Based on the analyses of the radii, intrinsic densities, overlap with the $3\alpha$ wave functions, and monopole strengths,
I summarize the main structures of the four kinds of the $0^+$ bands obtained in the present calculation as follows.
\begin{enumerate}
\item[(i)] The first band is predominantly a mixture of the shell-model state and the compact $3\alpha$ cluster state,
and the spin-orbit force makes the attractions to these states.

\item[(ii)] The second band is the $3\alpha$ cluster state with the $^8$Be+$\alpha$ correlation.
  The $0^+_2$ state involves the $\alpha$-cluster breaking due to the spin-orbit force,
  while the $2^+_2$ and $4^+_2$ states are rather pure $3\alpha$-cluster state.

\item[(iii)] The third band is the linear-chain state of $3\alpha$ configuration with a slight bending.
  The $0^+_3$ state involves the $\alpha$-cluster breaking due to the spin-orbit force,
  and the $2^+_3$ and $4^+_3$ state are rather pure $3\alpha$-cluster states.
  
\item[(iv)] The fourth band can be interpreted as the breathing mode of the second band,
  with a large relative distance between $^8$Be and the $\alpha$ particle, which enhances the radii.
  The $\alpha$-cluster breaking has a negligible effect on this band, resulting in the strong monopole transitions
  of $2^+_2 \to 2^+_4$ and $2^+_4 \to 4^+_4$.
  The transition of $0^+_2 \to 0^+_4$ is also significant, but not large, due to the $\alpha$-cluster breaking in the $0^+_2$ state. 
\end{enumerate}


\subsection{Quadrupole moment}

\begin{table}[t]
  \caption{
    Spectroscopic quadrupole moments of the $2^+$ and $4^+$ states of $^{12}$C
    in comparison with NCSpM \cite{dreyfuss13} and experiments \cite{raju18,saiz23}.
    Units are in $e\,$fm$^2$.
  }\vspace*{0.1cm}
\label{tab:Q}
\renewcommand{\arraystretch}{1.5}
\begin{tabular}{llrrrrlrrrrrr}
\hline  \vspace*{-0.10cm}
        && Present  && NCSpM    && Experiment \\
\hline                                                                   
 $2^+_1$~&& 6.08     && 5.9(1)   && $7.1\pm 2.5$,~~$9.3^{+3.5}_{-3.8}$ \\
 $2^+_2$~&& $-27.84$ && $-21(1)$ && \\[0.1cm]
 $4^+_1$~&& 6.33     &&  8.0(3)  && \\
 $4^+_2$~&& $-44.41$ && $-26(1)$ && \\
\hline
\end{tabular}
\end{table}

I discuss the deformation properties of $^{12}$C in terms of the quadrupole moments.
Table~\ref{tab:Q} shows the spectroscopic quadrupole moments of the $2^+_{1,2}$ and $4^+_{1,2}$ states of $^{12}$C.
I compare these values with the no-core symplectic shell model (NCSpM) calculations \cite{dreyfuss13} and the experimental results \cite{raju18,saiz23}.
My results are similar to the NCSpM results for four states.
For the $2^+_1$ state, my result is close to the experimental values within the margin of error.

It is found that the $2^+_1$ and $4^+_1$ states, which are considered to have a shell-model component,
commonly produce the positive values with similar magnitudes.
This indicates the oblate deformation of $^{12}$C in the intrinsic frame.
The $2^+_2$ and $4^+_2$ states commonly exhibit very large negative values, indicating the prolate deformation.
This originates from the developed $3\alpha$ clustering, which is confirmed by the intrinsic densities in Fig. \ref{fig:density5}.

\subsection{Quadrupole transition}

\begin{table}[t]
  \caption{
    Electric quadrupole transition strengths $B(E2)$ of $^{12}$C from the $2^+$ states in the multicool calculation
    in comparison with the experiments \cite{kelley17,zimmerman13a,zimmerman13b,dalessio20} and theories \cite{cheng25,suhara15,kanada07}.
    Units are in $e^2 {\rm fm}^4$.
  }\vspace*{0.1cm}
\label{tab:E2_2}
\renewcommand{\arraystretch}{1.5}
\begin{tabular}{clllllllllllllrrr}
\hline
                 && Expt        &  Present   & $3\alpha$(CNN) & $3\alpha$+$p_{3/2}$ & AMD \\
\hline               
$2^+_1\to 0^+_1$ && $7.63(19)$  &  8.64      &  8.9       & 7.4               & 8.5 \\
$2^+_1\to 0^+_2$ && $2.70(28)$  &  2.00      &            & 5.1               & 5.1 \\
$2^+_1\to 0^+_3$ &&  --         &  0.55      &            & 0.2               &     \\
$2^+_1\to 0^+_4$ &&  --         &  0.002     &            &                   &     \\[0.1cm]
$2^+_2\to 0^+_1$ && $0.73(13)$  &  1.03      & 0.9        & 1.1               &     \\
                 && $1.57^{+0.14}_{-0.11}$ & &            &                   &     \\
$2^+_2\to 0^+_2$ &&             &  233       &  78.2      & 76.5              & 100 \\
$2^+_2\to 0^+_3$ &&             &  182       & 113        & 166               & 310 \\[0.1cm]
$2^+_2\to 0^+_4$ &&             &  0.85      &            &                   &     \\
$2^+_3\to 0^+_1$ &&             &  0.02      &  0.01      &                   &     \\
$2^+_3\to 0^+_2$ &&             &  2.12      &  8.3       &                   & 6.4 \\
$2^+_3\to 0^+_3$ &&             &  75.3      & 88.8       &                   & 76  \\
$2^+_3\to 0^+_4$ &&             &  108       & 160        &                   &     \\[0.1cm]
$2^+_4\to 0^+_1$ &&             &  0.51      & 0.6        &                   &     \\
$2^+_4\to 0^+_2$ &&             & 11.8       & 10.7       &                   &     \\
$2^+_4\to 0^+_3$ &&             &  258       & 146        &                   &     \\
$2^+_4\to 0^+_4$ &&             &  196       & 809        &                   &     \\
\hline
\end{tabular}
\end{table}
\begin{table}[th]
  \caption{
    Electric quadrupole transition strengths $B(E2)$ of $^{12}$C from the $4^+$ states in the multicool calculation
    in comparison with other theories \cite{cheng25,kanada07}.
    Units are in $e^2 {\rm fm}^4$.
  }\vspace*{0.1cm}
\label{tab:E2_4}
\renewcommand{\arraystretch}{1.5}
\begin{tabular}{clllllllllllllrrr}
\hline
                 &&   Present & $3\alpha$(CNN) & AMD \\
\hline               
$4^+_1\to 2^+_1$ &&   12.8    & 10.9      & 16  \\
$4^+_1\to 2^+_2$ &&   24.1    &           & 7.5 \\
$4^+_1\to 2^+_3$ &&    0.46   &           &     \\
$4^+_1\to 2^+_4$ &&    0.42   &           &     \\[0.1cm]
$4^+_2\to 2^+_1$ &&    0.18   &   0.2     &     \\
$4^+_2\to 2^+_2$ &&   532     & 511       & 600 \\
$4^+_2\to 2^+_3$ &&   41.4    &  35.2     &  74 \\
$4^+_2\to 2^+_4$ &&   145     & 1202      &     \\[0.1cm]
$4^+_3\to 2^+_1$ &&   0.06    &   0.3     &    \\
$4^+_3\to 2^+_2$ &&   1.15    &  13.6     &    \\
$4^+_3\to 2^+_3$ &&   229     & 1633      &    \\
$4^+_3\to 2^+_4$ &&   272     & 533       &    \\[0.1cm]
$4^+_4\to 2^+_1$ &&   0.36    &  0.5      &    \\
$4^+_4\to 2^+_2$ &&   15.6    & 17.7      &    \\
$4^+_4\to 2^+_3$ &&   83.6    & 677       &    \\
$4^+_4\to 2^+_4$ &&   498     & 256       &    \\
\hline
\end{tabular}
\end{table}
%
I discuss the quadrupole transitions of $^{12}$C. In Tables \ref{tab:E2_2} and \ref{tab:E2_4},
I summarize the electric quadrupole transition strengths, $B(E2)$, associated with the $2^+$ states and the $4^+$ states, respectively,
including the experimental data \cite{kelley17,zimmerman13a,zimmerman13b,dalessio20} and other calculations \cite{cheng25,suhara15,kanada07}.
I find that my results are consistent with the experiments for three kinds of the $2^+ \to 0^+$ transitions.
The overall trend of the present results is the same as that of the other calculations \cite{cheng25,suhara15,kanada07} in the two tables.

In the $2^+ \to 0^+$ transitions of Table \ref{tab:E2_2}, the intraband transitions are basically strong in the four bands,
which is reasonable considering their internal structures.
In the interband transitions, some of which become strong, such as $2^+_2\to 0^+_3$, and $2^+_4\to 0^+_3$,
which exceed 100 $e^2$ fm$^4$. These strong transitions originate from the large radii of the $3\alpha$ cluster configurations
in the relevant states, as shown in Table \ref{tab:quanta}.
In the $4^+ \to 2^+$ transitions of Table \ref{tab:E2_4},
the intraband transitions are basically strong in the second, third, and fourth bands, which is consistent with other calculations \cite{cheng25,kanada07}.
This is also caused by the large radii of the relevant states in these band states.

For reference, I list the quadrupole transitions between the $2^+_{1-4}$ states and those between the $4^+_{1-4}$ states
in $^{12}$C in the appendix \ref{sec:appendix_A}.


\section{Summary}\label{sec:summary}

I investigated the structures of $^{12}$C, particularly the excited $0^+$ band states, which can have a $3\alpha$ cluster structure
that allows the $\alpha$-cluster breaking from the $s$-wave configuration.
In the microscopic nuclear wave function, I generated the optimal configurations using the antisymmetrized molecular dynamics (AMD);
the multiple AMD configurations are superposed and determined simultaneously in order to minimize the energy of the total system.
I further optimized the excited-state configurations by controlling their orthogonality to the ground-state configurations.

In AMD, the nucleon wave function is a Gaussian wave packet and
the centroid parameters of the Gaussians are determined using so-called the cooling method for the multiple AMD basis states as the energy minimization.
I therefore refer to this new framework as the multicool method.
In this paper, I applied the multicool method to $^{12}$C and focused on the structures of the band members of the $3\alpha$ cluster states.

The results show four kinds of the $0^+$ bands, which is consistent with the $3\alpha$ cluster model.
However, the order of the third and fourth bands differs from that predicted by the $3\alpha$ calculations.
This difference is due to the attractions of the spin-orbit force in the nucleon-nucleon interaction,
which can break the $\alpha$ clusters from the $s$-wave configuration in $^{12}$C,
and is not taken into account in the $3\alpha$ cluster model.

In the band states, the $0^+_1$ band is a mixture of the compact shell-model state and the compact $3\alpha$ cluster state
with relatively strong attractions of the spin-orbit force. 
The $0^+_2$ band is a spatially developed $3\alpha$ cluster state with the $^8$Be+$\alpha$ correlation.
The $0^+_2$ state partially involves the $\alpha$-cluster breaking with the coupling to the shell-model states,
which gains the attraction from the spin-orbit force.
On the other hand, the $2^+_2$ and $4^+_2$ states are rather pure $3\alpha$ cluster states.
The $0^+_3$ band is a linear-chain state with a slight bending, and the breaking of the $\alpha$ cluster occurs in the $0^+_3$ state,
while the $2^+_3$ and $4^+_3$ states are rather pure $3\alpha$ cluster states. This is the same result as for the $0^+_2$ band.
The $0^+_4$ band is considered a breathing mode of the $0^+_2$ band based on the analysis of the monopole transitions,
and this property agrees with the $3\alpha$ cluster calculations \cite{bo16,cheng25}.
In this band, the $3\alpha$ clusters are spatially separated and the breaking of the $\alpha$ cluster is negligible.

I discuss the charge form factors of the ground state of $^{12}$C.
The resulting distributions of the elastic and inelastic channels for the $0^+$ states are consistent with the experimental results.
The slight underestimation of the theoretical distributions in the higher momentum region beyond 2 fm$^{-1}$
might suggest the necessity of the tensor correlation.

I also evaluated the quadrupole moments and the quadrupole transitions of the $2^+$ and $4^+$ states of $^{12}$C.
My results reproduced the existing experimental values.
Regarding the quadrupole transitions, I confirm that the intraband transitions are often strong, and some of the interband transitions
exhibit large values. Transition strengths are often large in the $0^+_{2-4}$ bands that involve the spatially extended states.

In this study, the $0^+_{2-4}$ band states are treated in the bound-state approximation. 
In future, complex scaling is a promising method for describing these states as resonances above the $3\alpha$ threshold energy
with the correct boundary conditions \cite{myo14,zhang22,myo23a,kurokawa07}.

\section*{Acknowledgments}
The author would like to thank
Professor H. Horiuchi,
Professor H. Toki, 
Professor M. Lyu,
Professor Q. Zhao,
Professor N. Wan, 
Professor H. Takemoto,
Professor M. Isaka, 
Professor A. Dot\'e, and
Mr. J. Tian for valuable discussions.
This work was supported by JSPS KAKENHI Grants No. JP22K03643, No. JP25H01268, No. JP26K07092, and JST ERATO Grant No. JPMJER2304, Japan.
This work was also partly supported by the RCNP Collaboration Research Network program as the Project No. COREnet-059.
Numerical calculations were partly achieved through the use of the supercomputer system, SQUID, at the Cybermedia Center, Osaka University.

\appendix

\section{Quadrupole transition}\label{sec:appendix_A}
Table \ref{tab:E2_24} lists the electric quadrupole transition strengths of $^{12}$C between the $2^+_{1-4}$ states and between the $4^+_{1-4}$ states
for reference.

\begin{table}[th]
  \caption{
    Electric quadrupole transition strengths $B(E2)$ of $^{12}$C between the $2^+_{1-4}$ states and between the $4^+_{1-4}$ states
    in the multicool calculation.  Units are in $e^2 {\rm fm}^4$.
  }\vspace*{0.1cm}
\label{tab:E2_24}
\renewcommand{\arraystretch}{1.5}
\begin{tabular}{cccc|cccc}
\hline                
$2^+_1\to 2^+_2$ &&   0.14     &&&  $4^+_1\to 4^+_2$ &&  15.0    \\
$2^+_1\to 2^+_3$ &&   3.01     &&&  $4^+_1\to 4^+_3$ &&  0.97    \\
$2^+_1\to 2^+_4$ &&   0.10     &&&  $4^+_1\to 4^+_4$ &&  0.03    \\
$2^+_2\to 2^+_3$ &&   274      &&&  $4^+_2\to 4^+_3$ &&  44.1    \\
$2^+_2\to 2^+_4$ &&   20.7     &&&  $4^+_2\to 4^+_4$ &&  96.1    \\
$2^+_3\to 2^+_4$ &&   74.7     &&&  $4^+_3\to 4^+_4$ &&  16.9    \\
\hline
\end{tabular}
\end{table}

\bibliographystyle{apsrev4-1} 
\bibliography{reference_c12} 

\end{document}